\documentclass[smallcondensed,numbook,natbib]{svjour3}
\usepackage{float}
\usepackage{bbold}
\usepackage{physics}
\usepackage{multicol} 
\usepackage{makecell, multirow, tabularx} 
\usepackage[ruled,vlined]{algorithm2e}
\usepackage{bm}
\usepackage{soul}
\usepackage{siunitx}
\usepackage{adjustbox}
\usepackage{ulem}
\usepackage[title]{appendix}
\usepackage{algpseudocode}
\usepackage{graphicx}
\usepackage{amssymb}
\usepackage{booktabs}
\usepackage{mathrsfs}
\usepackage{bbm}
\usepackage{rotating}
\usepackage{lscape}
\usepackage{subfigure}
\usepackage{lipsum}
\usepackage{comment}
\usepackage[figurename=Fig.]{caption}
\usepackage{hyperref}
\hypersetup{colorlinks,linkcolor={blue},citecolor={blue},urlcolor={red}}
\usepackage[nameinlink,capitalise]{cleveref}
\newcommand{\ignore}[1]{}

\def\makeheadbox{{}}
\begin{document}

\title{A New Model-free Prediction Method: GA-NoVaS }
\author{Kejin Wu \and Sayar Karmakar}
\date{}
\institute {Kejin Wu \at University of California San Diego, La Jolla, CA 92093 \email{kwu@ucsd.edu} \and Sayar Karmakar \at University of Florida, Gainesville, FL 32611 \email{sayarkarmakar@ufl.edu} \and  Declarations of interest: none\\ \\ Corresponding author: Sayar Karmakar}

\maketitle

\begin{abstract}
Volatility forecasting plays an important role in the financial econometrics. Previous works in this regime are mainly based on applying various GARCH-type models. However, it is hard for people to choose a specific GARCH model which works for general cases and such traditional methods are unstable for dealing with high-volatile period or using small sample size. The newly proposed normalizing and variance stabilizing (NoVaS) method is a more robust and accurate prediction technique. This Model-free method is built by taking advantage of an inverse transformation which is based on the ARCH model. Inspired by the historic development of the ARCH to GARCH model, we propose a novel NoVaS-type method which exploits the GARCH model structure. By performing extensive data analysis, we find our model has better time-aggregated prediction performance than the current state-of-the-art NoVaS method on forecasting short and volatile data. The victory of our new method corroborates that and also opens up avenues where one can explore other NoVaS structures to improve on the existing ones or solve specific prediction problems.

\end{abstract}


\newpage
\section{Introduction}\label{sec:intro}
\noindent In the area of financial econometrics, forecasting volatility accurately and robustly is always important \citep{articleengle2001,du2007does}. Achieving high-quality volatility forecasting is crucial for practitioners and traders to make decisions in risk management, asset allocation, pricing of derivative instrument and strategic decisions of fiscal policies \citep{fang2018importance,ashiya2003directional, bansal2016risks,kitsul2013economics,morikawa2019uncertainty}. One example is that the global financial crisis of 2008 was severer than the predicted average recession by the US Congressional Budget Office. If people had a chance to perform a more accurate volatility prediction, this well-known crisis might not make huge negative impact on the global economics. A most recent case is that the COVID-19 pandemic. Some studies already warned the volatility of global financial market brought by the pandemic is even comparable with the financial crisis of 2008 \citep{abodunrin2020coronavirus,fernandes2020economic,yueh2020social}. Nobody knows when the next global uncertainty will come, thus a guideline for people to prepare for the latent financial recession is desired. Although we believe accurate economic predictions may derive people to make right and timely decisions which can potentially reduce effects created by financial recessions, volatility forecasting faces challenges of some facts, such as small sample size, heteroscedasticity and structural change \citep{chudy2020long}.

Standard methods to do volatility forecasting are typically built upon GARCH-type models, these models' abilities to forecast the absolute magnitude and quantiles or entire density of squared financial log-returns (i.e., equivalent to volatility forecasting to some extent)\footnote{Squared log-returns are an unbiased but very noisy measure of volatility pointed out by \citet{andersen1998answering}. Additionally, \citet{awartani2005predicting} showed that using squared log-returns as a proxy for volatility can render a correct ranking of different GARCH models in terms of a quadratic loss function.} were shown by \citet{articleengle2001} using Dow Jones Industrial Index. Later, many studies about comparing the performance of different GARCH-type models on predicting volatility of financial series were conducted \citep{chortareas2011forecasting,gonzalez2004forecasting,herrera2018forecasting,lim2013comparing,peters2001estimating,wilhelmsson2006garch,zheng2012empirical}. Some researchers also tried to develop the GARCH model further, such as adopting smoothing parameters or adding more related information to estimate model \citep{breitung2016simple,chen2012predicting,fiszeder2016low,taylor2004volatility}.  For modelling the proper process of volatility during fluctuated period, researchers also developed different approaches to achieve this goal, such as \citet{kim2011time} applied time series models with stable and tempered-stable innovations to measure market risk during highly volatile period; \citet{ben2014modelling} applied a
fractionally integrated time-varying GARCH (FITVGARCH) model to fit volatility; \citet{karmakar2020bayesian} developed a Bayesian method to estimate time-varying analogues of ARCH-type models for describing frequent volatility changes. Although there are many different types of GARCH models, it is a difficult problem to determine which single GARCH model outperforms others uniformly since the performance of these models heavily depends on the error distribution, the length of prediction horizon and the property of datasets. 

For getting out of this dilemma, we want to introduce a new idea to forecast volatility of financial series. Recall that during the prediction process, an inescapable intermediate process is building a model to describe historical data. Consequently, prediction results are restricted to this specific model used. However, the used model may not be correct within data, and the wrong model can even give better predictions sometimes \citep{politis2015modelfreepredictionprinciple}. Therefore, it is hard as a practitioner to determine which model should be used in this intermediate stage. With the Model-free Prediction Principle being first proposed by \citet{politis2003normalizing}, people can do predictions without the restriction of building a specific model and put all emphasis on data itself. Subsequently, we acquire a powerful method--NoVaS method--to do forecasting. 

The NoVaS method is one kind of Model-free method and applies the normalizing and variance-stabilizing transformation (NoVaS transformation) to do predictions \citep{politis2003normalizing,politis2007model,politis2015modelfreepredictionprinciple}. Some previous studies have shown that the NoVaS method possesses better predicting performance than GARCH-type models on forecasting squared log-returns, such as \citet{gulay2018comparison} showed that the NoVaS method could beat GARCH-type models (GARCH, EGARCH and GJR-GARCH) with generalized error distributions by comparing the pseudo-out of sample\footnote{The pseudo-out-of-sample forecasting analysis means using data up to and including current time to predict future values.} (POOS) forecasting performance. \citet{chen2020time} showed the ``Time-varying'' NoVaS method is robust against possible non-stationarities in the data. Furthermore, \citet{chen2019optimal} extended this NoVaS approach to do multi-step ahead predictions of squared log-returns. They also verified this approach could avoid error accumulation problem for single 5-steps ahead predictions and outperform the standard GARCH model for most of time. \cite{wu2021boosting} further substantiated the great performance of NoVaS methods on long-term (30-steps ahead) predictions. They considered taking the aggregated $h$-steps ahead prediction, i.e., the sliding-window mean of $h$-steps ahead predictions, to represent the long-term prediction. In the practical aspect of forecasting volatility, a single-point prediction may stand trivial meaning, since a small predicted volatility value at specific time point $t+h$ can not guarantee small values at other future time points. Conversely, this great-looking single prediction may mislead traders. Thus, a long-term prediction in the econometric area is meant to obtain some inferences about the future situation at overall level. In this article, we keep using the time-aggregated prediction value to measure different methods' short- long-term forecasting performance. This aggregated metric also had been applied to depict the future situation of electricity price or financial data \citep{chudy2020long, karmakar2020long,fryzlewicz2008normalized}.

To the best of our knowledge, previous studies in this regime mainly focus on comparing the NoVaS method with different GARCH-type models, and omit the potential of developing the NoVaS method itself further. Inspired by the development of ARCH model \citep{engle1982autoregressive} to GARCH model \citep{bollerslev1986generalized}, this article attempts to build a novel NoVaS method which is derived from the GARCH(1,1) structure. Through extensive data analysis, we show that our methods bring significant improvement when the available data is short in length or is of more volatile nature. These are usually challenging forecasting exercise and thus our new method can open up some new directions.

The rest of this article is organized as follows. Details about the existing NoVaS method and motivations to propose a new method are explained in \cref{sec:novas}. In \cref{sec:method}, we propose a new NoVaS transformation approach, then a named GA-NoVaS method is created. According to the procedure of proposing the refined GE-NoVaS-without-$\beta$ method in \citep{wu2021boosting}, we also put forward a parsimonious variant of the GA-NoVaS method. Moreover, we connect this new parsimonious method with the GE-NoVaS-without-$\beta$ method. For comparing our new methods with the current state-of-the-art NoVaS method, POOS predictions on different simulated datasets are performed in \cref{sec:simu}. Additionally, in \cref{sec:real data}, we deploy extensive data analysis using various types of real-world datasets. In \cref{sec:comparisonofpredictive}, we exhibit some statistical test results to manifest the reasonability of our new methods. Finally, future work and discussions are presented in \cref{sec:conclusion}.

\section{NoVaS method}\label{sec:novas}
\noindent In this section, details about the NoVaS method are given. We first introduce the Model-free Prediction Principle which is the core idea behind the NoVaS method. Then, we present how the NoVaS transformation can be built from an ARCH model. Finally, we give the motivation to build a new NoVaS transformation.

\subsection{Model-free prediction principle}\label{ssec:modelfreeprinciple}
\noindent Before presenting the NoVaS method in detail, we throw some light on the insight of the Model-free Prediction Principle. The main idea behind this is applying an invertible transformation function $H_n$ which can map a non-$i.i.d$. vector $\{Y_i~;i = 1,\cdots,n\}$ to a vector $\{\epsilon_i;~i=1,\cdots,n\}$ that has $i.i.d.$ components. Since the prediction of $i.i.d.$ data is somewhat standard, the prediction of $Y_{n+1}$ can be easily obtained by inversely transforming $\hat{\epsilon}_{n+1}$ which is a prediction of $\epsilon_{n+1}$ using $H_n^{-1}$. For example, we can express the prediction $\hat{Y}_{n+1}$ as a function of $\bm{Y}_n$, $\bm{X}_{n+1}$ and $\hat{\epsilon}_{n+1}$:
\begin{equation}
     \hat{Y}_{n+1}=f_{n+1}(\bm{Y}_{n}, \bm{X}_{n+1},\hat{\epsilon}_{n+1}) \label{3.1e1}
\end{equation}
Where, $\bm{Y}_{n}$ denotes all historical data $\{Y_t;~i =1,\cdots,n\}$; $\bm{X}_{n+1}$ is the collection of all predictors and it also contains the value of future predictor $X_{n+1}$. Although a qualified transformation $H_n$ is hard to find for general case, we have naturally existing forms of $H_n$ in some situations, such as in linear time series environment. In this article, we will show how to utilize existing forms--ARCH and GARCH models--to build NoVaS transformations to predict volatility. One thing should be noticed is that a more complicated procedure to obtain transformation $H_n$ is needed (see \citep{politis2015modelfreepredictionprinciple} for more details) if we do not have these existing forms for some situations. 

After acquiring \cref{3.1e1}, we can even predict the general form of $Y_{n+1}$ such as $g(Y_{n+1})$. \citet{politis2015modelfreepredictionprinciple} defined two data-based optimal predictors of $g(Y_{n+1})$  under $L_1$ (Mean Absolute Deviation) and $L_2$ (Mean Squared Error) criterions respectively as below:
\begin{equation}
\begin{split}
    g(Y_{n+1})_{L_2} &= \frac{1}{M}\sum_{m=1}^Mg(f_{n+1}(\bm{Y}_n,\bm{X}_{n+1},\hat{\epsilon}_{n+1,m}))\\
    g(Y_{n+1})_{L_1} &= \text{Median~of~}\{g(f_{n+1}(\bm{Y}_n,\bm{X}_{n+1},\hat{\epsilon}_{n+1,m}))|m = 1,\cdots,M\} \label{3.1e2}
    \end{split}
\end{equation}
In \cref{3.1e2}, $\{\hat{\epsilon}_{n+1,m}\}_{m=1}^{M}$ are generated from its own distribution by bootstrapping or from a normal distribution through Monte Carlo method (recall the $\{\epsilon_i;~i=1,\cdots,n\}$ are $i.i.d.$); $M$ takes a large number (5000 in this article). 

\subsection{NoVaS transformation}\label{novastransformation}
\noindent The NoVaS transformation is a straightforward application of the Model-free Prediction Principle. More specifically, the NoVaS transformation is a qualified function $H_n$ which is based on the ARCH model introduced by \citet{engle1982autoregressive} as follows:
\begin{equation}
    Y_t = W_t\sqrt{a+\sum_{i=1}^pa_iY_{t-i}^2} \label{3.2e1}
\end{equation}
In \cref{3.2e1}, these parameters satisfy $a\geq 0$, $a_i\geq 0$, for all $i = 1,\cdots,p$; $W_t\sim i.i.d.~N(0,1)$. In other words, the structure of the ARCH model gives us a ready-made $H_n$. We can express $W_t$ in \cref{3.2e1} using other terms:
\begin{equation}
    W_t = \frac{Y_t}{\sqrt{a+\sum_{i=1}^pa_iY_{t-i}^2}} ~;~\text{for}~ t=p+1,\cdots,n. \label{3.2e2}
\end{equation}
 Subsequently, \cref{3.2e2} can be seen as a potential form of $H_n$. However, some additional adjustments were made by \citet{politis2003normalizing}. Firstly, $Y_t$ was added into the denominator on the right hand side of \cref{3.2e2} for obeying the rule of causal estimate which means only using present and past data, and utilizing as much information as possible. Secondly, the constant $a$ was replaced by $\alpha s_{t-1}^2$ to create a scale invariant parameter $\alpha$. Consequently, after observing $\{Y_1,\cdots,Y_n\}$, the adjusted \cref{3.2e2} can be written as \cref{3.2e3}:
\begin{equation}
      W_{t}=\frac{Y_t}{\sqrt{\alpha s_{t-1}^2+\beta Y_t^2+\sum_{i=1}^pa_iY_{t-i}^2}}~;~\text{for}~ t=p+1,\cdots,n. \label{3.2e3}
\end{equation}
In \cref{3.2e3}, $\{Y_t;~t=1,\cdots,n\}$ is the target data, such as financial log-returns in this paper; $\{W_{t};~t=p+1,\cdots,n\}$ is the transformed vector; $\alpha$ is a fixed scale invariant constant; $s_{t-1}^2$ is an estimator of the variance of $\{Y_i;~i = 1,\cdots,t-1\}$ and can be calculated by $(t-1)^{-1}\sum_{i=1}^{t-1}(Y_i-\mu)^2$, with $\mu$ being the mean of $\{Y_i;~i = 1,\cdots,t-1\}$.

$\{W_t;~t=p+1,\cdots,n\}$ expressed in \cref{3.2e3} are assumed to be $i.i.d.~N(0,1)$, but it is not true yet. For making \cref{3.2e3} be a qualified function $H_n$ (i.e., making $\{W_t\}_{t=p+1}^{n}$ really obey standard normal distribution), we still need to add some restrictions on $\alpha$ and $\beta, a_1,\cdots,a_p$. Recalling the NoVaS transformation means normalizing and variance stabilizing transformation, we first stabilize the variance by requiring:
\begin{equation}
    \alpha\geq0, \beta\geq0, a_i\geq0~;~\text{for all}~i\geq1, \alpha + \beta + \sum_{i=1}^pa_i=1. \label{3.2e4}
\end{equation} 
By requiring unknown parameters in \cref{3.2e3} to satisfy \cref{3.2e4}, we can make $\{W_t\}_{t=p+1}^{n}$ series possess approximate unit variance. Additionally, $\alpha$ and $\beta,a_1,\cdots,a_p$ must be selected carefully. In the work of \citet{politis2015modelfreepredictionprinciple}, two different structures of $\beta, a_1,\cdots,a_p$ were provided:
\begin{equation}
\begin{split}
    &\text{Simple NoVaS:}~\alpha =0, \beta = \frac{1}{p+1}, a_i = \frac{1}{p+1}~;~\text{for all}~1\leq i\leq p.\\
    &\text{Exponential NoVaS:}~\alpha =0, \beta = c' ,a_i = c'e^{-ci}~;~\text{for all}~1\leq i\leq p,~c' = \frac{1}{\sum_{j=0}^pe^{-cj}}. \label{3.2e5}
    \end{split}
\end{equation}
We keep using S-NoVaS and E-NoVaS to denote these two NoVaS methods in \cref{3.2e5}. For the S-NoVaS, all $\beta,a_1,\cdots,a_p$ taking same value means we assign same weights on past data. Similarly, for the E-NoVaS, $\beta,a_1,\cdots,a_p$ are exponentially positive decayed coefficients, which means we assign decayed weights on the past data. Note that $\alpha$ is equal to 0 in both methods above. If we allow $\alpha$ takes a positive small value, we can get two different methods:
\begin{equation}
\begin{split}
    &\text{Generalized Simple NoVaS:}~\alpha \neq 0, \beta =  \frac{1-\alpha}{p+1}, a_i = \frac{1-\alpha}{p+1}~;~\text{for all}~1\leq i\leq p.\\
    &\text{Generalized Exponential NoVaS:}~\alpha \neq 0, \beta = c', a_i = c'e^{-ci}~;~\text{for all}~1\leq i\leq p,\\
    &\text{where, }c' = \frac{1-\alpha}{\sum_{j=0}^pe^{-cj}}. \label{3.2e6}
    \end{split}
\end{equation}
We keep using GS-NoVaS and GE-NoVaS to denote these two generalized NoVaS methods in \cref{3.2e6}. $\alpha$ in both generalized methods takes value from $(0,1)$\footnote{If $\alpha = 1$, all $a_i$ will equal to 0. It means we just standardize $\{Y_t\}$ and do not utilize the structure of ARCH model.}. Obviously, NoVaS and generalized NoVaS methods all meet the requirement of \cref{3.2e4}. 

Finally, we still need to make $\{W_t\}_{t=p+1}^{n}$ independent. In practice, $\{W_t\}_{t=p+1}^{n}$ transformed from financial log-returns by NoVaS transformation are usually uncorrelated\footnote{If $\{W_t\}_{t=p+1}^{n}$ is correlated, some additional adjustments need to be done, more details can be found in \citep{politis2015modelfreepredictionprinciple}.}. Therefore, if we make the empirical distribution of $\{W_t\}_{t=p+1}^{n}$ close to the standard normal distribution (i.e., normalizing $\{W_t\}_{t=p+1}^{n}$), we can get a $i.i.d.$ series $\{W_t\}_{t=p+1}^{n}$. Note that the distribution of financial log-returns is usually symmetric, thus, the kurtosis can serve as a simple distance measuring the departure of a non-skewed dataset from the standard normal distribution whose kurtosis is 3 \citep{politis2015modelfreepredictionprinciple}. We use $\hat{F}_w$ to denote the empirical distribution of $\{W_t\}_{t=p+1}^{n}$ and use $KURT(W_t)$ to denote the kurtosis of $\hat{F}_w$. Then, for making $\hat{F}_w$ close to the standard normal distribution, we attempt to minimize $\abs{KURT(W_t)-3}$\footnote{More details about this minimizing process can be found in \cite{politis2015modelfreepredictionprinciple}.} to obtain the optimal combination of $\alpha,\beta,a_1,\cdots,a_p$. Consequently, the NoVaS transformation is determined.

From the thesis of \citet{chen2018prediction}, Generalized NoVaS methods are better for interval prediction and estimation of squared log-returns than other NoVaS methods. The GE-NoVaS method which assigns exponentially decayed weight to the past data is also more reasonable than the GS-NoVaS method which handles past data equally. Therefore, in this article, we verify the advantage of our new methods by comparing them with the GE-NoVaS method. Before going further to propose the new NoVaS transformation, we talk more details about the GE-NoVaS method and our motivations to create new methods.

\subsection{GE-NoVaS method}\label{ssec:genovas}
\noindent For the GE-NoVaS method, the fixed $\alpha$ is larger than 0 and selected from a grid of possible values based on predictions performance. In this article, we define this grid as $(0.1,0.2,\cdots,0.8)$ which contains 8 discrete values\footnote{It is possible to refine this grid to get a better transformation. However, computation burden will also increase.}. From \cref{novastransformation}, based on the guide of Model-free Prediction Principle, we already get the function $H_n$ of GE-NoVaS method by requiring parameters to satisfy \cref{3.2e4} and minimizing $\abs{KURT(W_t)-3}$ . For completing the Model-free prediction process, we still need to figure out the form of $H_n^{-1}$. Through \cref{3.2e3}, $H_n^{-1}$ can be written as follows:
\begin{equation}
    Y_t=\sqrt{\frac{W_{t}^2}{1-\beta W_{t}^2}(\alpha s_{t-1}^2+\sum_{i=1}^pa_iY_{t-i}^2)}~;~\text{for}~ t=p+1,\cdots,n. \label{3.3e1}
\end{equation}
Based on \cref{3.3e1}, it is easy to get the analytical form of $Y_{n+1}$, which can be expressed as below:
\begin{equation}
    Y_{n+1}=\sqrt{\frac{W_{n+1}^2}{1-\beta W_{n+1}^2}(\alpha s_{n}^2+\sum_{i=1}^pa_iY_{n+1-i}^2)} \label{3e10}
\end{equation}
In \cref{3e10}, $s_n^2$ is an estimator of the variance of $\{Y_i;~i=1,\cdots,n\}$ and can be calculated by $n^{-1}\sum_{i=1}^n(Y_i-\mu)^2$, $\mu$ is the mean of data; $W_{n+1}$ will be replaced by a sample from the empirical distribution $\hat{F}_w$ or the trimmed standard normal distribution\footnote{The reason of utilizing a trimmed standard normal distribution is transformed $\{W_t\}_{t=p+1}^{n}$ are between $-1/\sqrt{\beta}$ and $1/\sqrt{\beta}$ from \cref{3.2e3}.}. Based on aforementioned $L_1$ and $L_2$ optimal predictor in \cref{3.1e2}, we can define $L_1$ and $L_2$ optimal predictors of $Y_{n+1}^2$ after observing historical information set $\mathscr{F}_{n} = \{Y_t,1\leq t \leq n\}$ as below:
\begin{equation}\label{Eqq.2.4}
\begin{split}
     L_1\text{-optimal predictor of}~&Y_{n+1}^2:\\ &\text{Median}\left\{Y_{n+1,m}^2:m=1,\cdots,M\big\rvert \mathscr{F}_{n}\right\} \\ 
     &= \text{Median}\left\{\frac{W_{n+1,m}^2}{1-\beta W_{n+1,m}^2}(\alpha s_{n}^2+\sum_{i=1}^pa_iY_{n+1-i}^2): m=1,\cdots,M \bigg\rvert \mathscr{F}_{n}\right\} \\
    &=(\alpha s_{n}^2+\sum_{i=1}^pa_iY_{n+1-i}^2)\text{Median}\left\{\frac{W_{n+1,m}^2}{1-\beta W_{n+1,m}^2}:m=1,\cdots,M\right\} \\
    L_2\text{-optimal predictor of}~&Y_{n+1}^2:\\
    &\text{Mean}\left\{Y_{n+1,m}^2:m=1,\cdots,M \big\rvert \mathscr{F}_{n}\right\} \\
    &= \text{Mean}\left\{\frac{W_{n+1,m}^2}{1-\beta W_{n+1,m}^2}(\alpha s_{n}^2+\sum_{i=1}^pa_iY_{n+1-i}^2):m=1,\cdots,M\bigg\rvert \mathscr{F}_{n}\right\} \\
    &=(\alpha s_{n}^2+\sum_{i=1}^pa_iY_{n+1-i}^2)\text{Mean}\left\{\frac{W_{n+1,m}^2}{1-\beta W_{n+1,m}^2}:m=1,\cdots,M\right\} 
    \end{split}
\end{equation}
where, $\{W_{n+1,m}\}_{m=1}^{M}$ are bootstrapped $M$ times from its empirical distribution or generated from a trimmed standard normal distribution by Monte Carlo method. In other words, $Y_{n+1}$ can be presented as a function of $W_{n+1}$ and $\mathscr{F}_{n}$ as below:
\begin{equation}
    Y_{n+1} = f_{GE}(W_{n+1};\mathscr{F}_{n}) \label{3e11}
\end{equation}
For reminding us this relationship between $Y_{n+1}$ and $W_{n+1}, Y_1, \cdots, Y_n$ is derived from the GE-NoVaS method, we use $f_{GE}(\cdot)$ to denote this function. It is not hard to find $Y_{n+2}$ can be expressed as:
\begin{equation}
\begin{split}
   Y_{n+2} &= \sqrt{\frac{W_{n+2}^2}{1-\beta W_{n+2}^2}(\alpha s_{n}^2+\sum_{i=1}^pa_iY_{n+2-i}^2)}\\
  &=f_{GE}(W_{n+1},W_{n+2};\mathscr{F}_{n}) \label{3e12}
\end{split}
\end{equation}
For deriving the optimal predictor of $Y_{n+2}^{2}$ we can generate $\{W_{n+1,m},W_{n+2,m}\}_{m=1}^{M}$ $M$ times to compute the $L_1$ and $L_2$ optimal predictors of $Y_{n+1}^{2}$ firstly. Then, with predicted $\hat{Y}_{n+1}^{2}$ and $\{W_{n+2,m}\}_{m=1}^{M}$, $L_1$ and $L_2$ optimal predictors of $Y_{n+2}^{2}$ can be obtained similarly with \cref{3e12}. 

Finally, iterating the process to get predicted $\hat{Y}_{n+2}^{2}$, we can accomplish the multi-step ahead prediction of $Y_{n+h}^{2}$ for any $h\geq 3$. Basically speaking, $\{W_{n+1,m},\cdots,$ $W_{n+h,m}\}_{m=1}^{M}$ are needed. Then, we iteratively predict $Y_{n+1}^{2}$, $Y_{n+2}^{2}$, $Y_{n+3}^{2}$ and so on. In summary, we can express $Y_{n+h}$ as:

\begin{equation}
    Y_{n+h} = f_{GE}(W_{n+1},\cdots,W_{n+h};\mathscr{F}_{n})~;~\text{for any}~h\geq1. \label{3e13}
\end{equation}
Note that, the analytical form of $Y_{n+h}$ from the GE-NoVaS transformation, only depends on $i.i.d.~\{W_{n+1},\cdots,W_{n+h}\}$ and $\mathscr{F}_{n}$.

\subsection{Motivations of building a new NoVaS transformation}\label{ssecmotivation}
\noindent 

\textit{Structured form of coefficients:} In last few subsections, we illustrated the procedure of using the GE-NoVaS method to calculate $L_1$ and $L_2$ predictors of $Y_{n+h}$. However, the form of coefficients $\beta, a_1,\cdots,a_p$ in \cref{3.2e3} is somewhat arbitrary. Note that the GE-NoVaS method simply sets $\beta, a_1,\cdots,a_p$ to be exponentially decayed. This lets us put forward the following idea: Can we build a more rigorous form of $\beta, a_1,\cdots,a_p$ based on the relevant model itself without assigning any prior form on coefficients? In this paper, a new approach to explore the form of $\beta, a_1,\cdots,a_p$ based on the GARCH(1,1) model is proposed. Subsequently, the GARCH-NoVaS (GA-NoVaS) transformation is built.

\textit{Changing the NoVaS transformation:} The authors in \citet{chen2019optimal} claimed that NoVaS method can generally avoid the error accumulation problem which is derived from the traditional multi-stage prediction, i.e., using predicted values to predict further data iteratively. However, \cite{wu2021boosting} showed the current state-of-the-art GE-NoVaS method still renders extremely large time-aggregated multi-step ahead predictions under $L_2$ risk measure sometimes. The reason for this phenomenon is the denominator of \cref{3e10} will be quite small when the generated $W^*$ is very close to $\sqrt{1/\beta}$. In this situation, the prediction error will be amplified. Moreover, when the long-step ahead prediction is desired, this amplification will be accumulated and the final prediction will be ruined. Thus, a $\beta$-removing technique was imposed on the GE-NoVaS method to get a GE-NoVaS-without-$\beta$ method\footnote{In the paper of \citep{wu2021boosting}, they used $a_0$ to be the coefficient of $Y_t^2$, so they defined a $a_0$-removing technique Here, we use $\beta$.}. Similarly, we can also obtain a parsimonious variant of the GA-NoVaS method by reusing this technique. The discussion of these parsimonious methods is presented in \cref{subsec:parsimoniousvariant,ssec:connection}.

\section{Our new methods}\label{sec:method}
\noindent In this section, we first propose the GA-NoVaS method which is directly developed from the GARCH(1,1) model without assigning any specific form of $\beta, a_1,\cdots,a_p$. Then, the GA-NoVaS-without-$\beta$ method is introduced through applying the $\beta$-removing technique again. We also provide algorithms of these two new methods in the end. 

\subsection{GA-NoVaS transformation}\label{ssc:garchnovas}
\noindent Recall the GE-NoVaS method mentioned in \cref{ssec:genovas}, it was built by taking advantage of the ARCH(p) model, $p$ takes an initially large value in the algorithm of the GE-NoVaS method. Although the ARCH model is the base of the GE-NoVaS method, we should notice that free parameters of the GE-NoVaS method are just $c$ and $\alpha$. For representing $p$ coefficients by just two free parameters, some specific forms are assigned to $\beta, a_1,\cdots,a_p$. However, we doubt the reasonability of this approach. Thus, we try to use a more convincing approach to find $\beta, a_1,\cdots,a_p$ directly without assigning any prior form to these parameters. We call this NoVaS transformation method by GA-NoVaS. 

The idea behind this new method is inspired by the successful historic development of the ARCH to GARCH model. The popular GARCH(1,1) model can be utilized to represent the ARCH($\infty$) model, the proof is trivial, see \citep{politis2019time} for some references. If we want to build a GE-NoVaS transformation with $p$ converging to $\infty$, it is appropriate to replace the denominator at the right hand side of \cref{3.2e2} by the structure of GARCH(1,1) model. We present the GARCH(1,1) model as \cref{4e1}:
\begin{equation}
    \begin{split}
        Y_t &= \sigma_tW_t\\
        \sigma_t^2&=a + a_1Y_{t-1}^2 + b_1\sigma_{t-1}^2 \label{4e1}
    \end{split}
\end{equation}
In \cref{4e1}, $a \geq 0$, $a_1 > 0$, $b_1 > 0$, and $W_t\sim i.i.d.~N(0,1)$. In other words, a potentially qualified transformation related to the GARCH(1,1) or ARCH($\infty$) model can be exhibited as:
\begin{equation}
    W_t = \frac{Y_t}{\sqrt{a + a_1Y_{t-1}^2 + b_1\sigma_{t-1}^2}} \label{eq:3.2}
\end{equation}
However, recall the core insight of the NoVaS method is connecting the original data with the transformed data by a qualified transformation function. A primary problem is desired to be solved is that the right-hand side of \cref{eq:3.2} contains other terms rather than only $\{Y_t\}$ terms. Thus, more manipulations are required to build the GA-NoVaS method. Taking \cref{4e1} as the starting point, we first find out expressions of $\sigma_{t-1}^2,\sigma_{t-2}^2,\cdots$ as follow:
\begin{equation}
    \begin{split}
        \sigma_{t-1}^2 &= a + a_1Y_{t-2}^2 + b_1\sigma_{t-2}^2\\
        \sigma_{t-2}^2 &= a + a_1Y_{t-3}^2 + b_1\sigma_{t-3}^2\\
        \vdots& \label{4e2}
    \end{split}
\end{equation}
Plug all components in \cref{4e2} into \cref{4e1}, one equation sequence can be gotten:
\begin{equation}
    \begin{split}
        Y_t &= W_t\sqrt{a + a_1Y_{t-1}^2 + b_1\sigma_{t-1}^2}\\ 
        &= W_t\sqrt{a + a_1Y_{t-1}^2 + b_1(a + a_1Y_{t-2}^2 + b_1\sigma_{t-2}^2)}\\
        &= W_t\sqrt{a + a_1Y_{t-1}^2 + b_1a + b_1a_1Y_{t-2}^2 + b_1^2(a + a_1Y_{t-3}^2 + b_1\sigma_{t-3}^2)}\\
        &\vdots \label{4e3}
    \end{split}
\end{equation}
Iterating the process in \cref{4e3}, with the requirement of $a_1+b_1<1$ for the stationarity, the limiting form of $Y_t$ can be written as \cref{4e4}:
\begin{equation}
   Y_t =W_t\sqrt{ \sum_{i = 1}^{\infty}a_1b_1^{i-1}Y_{t-i}^2 + \sum_{j=0}^{\infty}ab_1^j} = W_t\sqrt{ \sum_{i = 1}^{\infty}a_1b_1^{i-1}Y_{t-i}^2 + \frac{a}{1-b_1}}   \label{4e4}
\end{equation}
We can rewrite \cref{4e4} to get a potential function $H_n$ which is corresponding to the GA-NoVaS method:
\begin{equation}
    W_t = \frac{Y_t}{\sqrt{ \sum_{i = 1}^{\infty}a_1b_1^{i-1}Y_{t-i}^2 + \frac{a}{1-b_1}}} \label{4e5}
\end{equation}
Recall the adjustment taken in the existing GE-NoVaS method, the total difference between \cref{3.2e2,3.2e3} can be seen as the term $a$ being replaced by $\alpha s_{t-1}^2 + \beta Y_t^2$. Apply this same adjustment on \cref{4e5}, then this equation will be changed to the form as follows:
\begin{equation}
    W_t = \frac{Y_t}{\sqrt{ \frac{\beta Y_t^2 + \alpha s_{t-1}^2}{1-b_1}+ \sum_{i = 1}^{\infty}a_1b_1^{i-1}Y_{t-i}^2 }} =  \frac{Y_t}{\sqrt{ \frac{\beta Y_t^2}{1-b_1}+ \frac{\alpha s_{t-1}^2}{1-b_1} + \sum_{i = 1}^{\infty}a_1b_1^{i-1}Y_{t-i}^2 }} \label{4e6}
\end{equation}
 In \cref{4e6}, since $\alpha/(1-b_1)$ is also required to take a small positive value, this term can be seen as a $\Tilde{\alpha}$ ($\Tilde{\alpha} \geq 0$) which is equivalent with $\alpha$ in the existing GE-NoVaS method. Thus, we can simplify $\alpha s_{t-1}^2/(1-b_1)$ to $\Tilde{\alpha} s_{t-1}^2$. For keeping the same notation style with the GE-NoVaS method, we use $\alpha s_{t-1}^2$ to represent $\alpha s_{t-1}^2/(1-b_1)$. Then \cref{4e6} can be represented as:
 \begin{equation}
    W_t =  \frac{Y_t}{\sqrt{ \frac{\beta Y_t^2}{1-b_1}+ \alpha s_{t-1}^2 + \sum_{i = 1}^{\infty}a_1b_1^{i-1}Y_{t-i}^2 }} \label{4e7}
\end{equation}
 For getting a qualified GA-NoVaS transformation, we still need to make the transformation function \cref{4e7} satisfy the requirement of the Model-free Prediction Principle. Recall that in the existing GE-NoVaS method, $\alpha + \beta + \sum_{i=1}^pa_i$ in \cref{3.2e3} is restricted to be 1 for meeting the requirement of variance-stabilizing and the optimal combination of $\alpha,\beta, a_1,\cdots,a_p$ is selected to make the empirical distribution of $\{W_t\}$ as close to the standard normal distribution as possible (i.e., minimizing $\abs{KURT(W_t)-3}$). Similarly, for getting a qualified $H_n$ from \cref{4e7}, we require:
 \begin{equation}
     \frac{\beta}{1-b_1} +\alpha + \sum_{i=1}^{\infty}a_1b_1^{i-1} = 1 \label{4e8}
 \end{equation}
  Under this requirement, since $a_1$ and $b_1$ are both less than 1, $a_1b_1^{i-1}$ will converge to 0 as $i$ converges to $\infty$, i.e., $a_1b_1^{i-1}$ is neglectable when $i$ takes large values. So it is reasonable to replace $\sum_{i=1}^{\infty}a_1b_1^{i-1}$ in \cref{4e8} by $\sum_{i=1}^{q}a_1b_1^{i-1}$, where $q$ takes a large value. Then a truncated form of \cref{4e7} can be written as \cref{4e9}:
\begin{equation}
    W_t = \frac{Y_t}{\sqrt{ \frac{\beta Y_t^2}{1-b_1}+ \alpha s_{t-1}^2 + \sum_{i = 1}^{q}a_1b_1^{i-1}Y_{t-i}^2 }}~;~\text{for}~ t=q+1,\cdots,n. \label{4e9}
\end{equation}
Now, we take \cref{4e9} as a potential function $H_n$. Then, the requirement of variance-stabilizing is changed to:
\begin{equation}
    \frac{\beta}{1-b_1} +\alpha + \sum_{i=1}^{q}a_1b_1^{i-1} = 1\label{4e10}
\end{equation}
\\
Akin to \cref{3.2e6}, we scale $\{\frac{\beta}{1-b_1},a_1,a_1b_1$ $,a_1b_1^{2},$ $\cdots,a_1b_1^{q-1} \}$ of \cref{4e10} by timing a scalar $\frac{1-\alpha}{\frac{\beta}{1-b_1} + \sum_{i=1}^{q}a_1b_1^{i-1}}$, and then search optimal coefficients. For presenting \cref{4e9} with scaling coefficients in a concise form, we use $\{c_0,c_1,\cdots,c_q\}$ to represent $\{\frac{\beta}{1-b_1},a_1,a_1b_1$ $,a_1b_1^{2},$ $\cdots,a_1b_1^{q-1} \}$ after scaling, which implies that we can rewrite \cref{4e9} as:
\begin{equation}
    W_t = \frac{Y_t}{\sqrt{ c_0Y_t^2+ \alpha s_{t-1}^2 + \sum_{i = 1}^{q}c_iY_{t-i}^2 }}~;~\text{for}~ t=q+1,\cdots,n. \label{4e9v}
\end{equation}
\begin{remark}(The difference between GA-NoVaS and GE-NoVaS methods)
Compared with the existing GE-NoVaS method, we should notice that the GA-NoVaS method possesses a totally different transformation structure. Recall all coefficients except $\alpha$ implied by the GE-NoVaS method are expressed as $\beta = c', a_i = c'e^{-ci}~$ $\text{for all}~1\leq$ $i\leq p$, $c' = \frac{1-\alpha}{\sum_{j=0}^pe^{-cj}}$. There are only two free parameters $c$ and $\alpha$. However, there are four free parameters $\beta, a_1, b_1$ and $\alpha$ in \cref{4e9}. For example, the coefficient of $Y_t^2$ of the GE-NoVaS method is $(1-\alpha)/(\sum_{j=0}^pe^{-cj})$. On the other hand, the corresponding coefficient in the GA-NoVaS structure is $c_0 = \beta(1-\alpha)/(\beta+(1-b_1)\sum_{i=1}^{q}a_1b_1^{i-1})$. We can think the freedom of coefficients within the GA-NoVaS is larger than the freedom in the GE-NoVaS. At the same time, the structure of GA-NoVaS method is built from GARCH(1,1) model directly without imposing any prior assumption on coefficients. We believe this is the reason why our GA-NoVaS method shows better prediction performance in \cref{sec:simu,sec:real data}. 
\end{remark}

Furthermore, for achieving the aim of normalizing, we still fix $\alpha$ to be one specific value from $\{0.1,0.2,\cdots,0.8\}$, and then search the optimal combination of $\beta,a_1,b_1$ from three grids of possible values of $\beta,a_1,b_1$ to minimize $\abs{KURT(W_t)-3}$. After getting a qualified $H_n$, $H_n^{-1}$ will be outlined immediately:
\begin{equation}
    Y_t = \sqrt{\frac{W_t^2}{1-c_0W_t^2}(\alpha s_{t-1}^2+\sum_{i=1}^qc_iY_{t-i}^2)}~;~\text{for}~ t=q+1,\cdots,n. \label{4e11}
\end{equation}
Based on \cref{4e11}, $Y_{n+1}$ can be expressed as the equation follows:
\begin{equation}
    Y_{n+1} = \sqrt{\frac{W_{n+1}^2}{1-c_0W_{n+1}^2}(\alpha s_{n}^2+\sum_{i=1}^qc_iY_{n+1-i}^2)}  \label{4e12}
\end{equation}
%\begin{equation}
%    Y_{n+1} = \sqrt{\frac{W_{n+1}^2(1-b_1)(\alpha s_{n}^2+\sum_{i=1}^qa_1b_1^{i-1}Y_{n+1-i}^2)}{1-b_1-W_{n+1}^2\beta}}\label{4e12}
%\end{equation}
Also, it is not hard to express $Y_{n+h}$ as a function of $W_{n+1},\cdots, W_{n+h} $ and $\mathscr{F}_{n}$ with GA-NoVaS method like we did in \cref{ssec:genovas}:
\begin{equation}
    Y_{n+h} = f_{GA}(W_{n+1},\cdots,W_{n+h};\mathscr{F}_{n})~;~\text{for any}~h\geq 1. \label{4e13}
\end{equation}
\par
\noindent Once the expression of $Y_{n+h}$ is figured out, we can apply the same procedure with the GE-NoVaS method to get the optimal predictor of $Y_{n+h}$ under $L_1$ or $L_2$ risk criterion. To deal with $\alpha$, we still adopt the same strategy used in the GE-NoVaS method, i.e., select the optimal $\alpha$ from a grid of possible values based on prediction performance. One thing should be noticed is that the value of $\alpha$ is invariant during the process of optimization once we fix it as a specific value. More details about the algorithm of this new method can be found in \cref{ssc:algorithm}.

\subsection{Parsimonious variant of the GA-NoVaS method}\label{subsec:parsimoniousvariant}
According to the $\beta$-removing idea, we can continue proposing the GA-NoVaS-without-$\beta$ method which is a parsimonious variant of the GA-NoVaS method. From \cite{wu2021boosting}, functions $H_n$ and $H_n^{-1}$ corresponding to the GE-NoVaS-without-$\beta$ method can be presented as follow:
\begin{equation}
     W_{t}=\frac{Y_t}{\sqrt{\alpha s_{t-1}^2+\sum_{i=1}^pa_iY_{t-i}^2}}~;~Y_t=\sqrt{W_{t}^2(\alpha s_{t-1}^2+\sum_{i=1}^pa_iY_{t-i}^2)}~;~\text{for}~ t=p+1,\cdots,n. \label{eq:3e15} 
\end{equation}
\cref{eq:3e15} still need to satisfy the requirement of normalizing and variance-stabilizing transformation. Therefore, we restrict $\alpha + \sum_{i=1}^pa_i = 1$  and still select the optimal combination of $ a_1,\cdots,a_p$ by minimizing $\abs{KURT(W_t)-3}$. Then, $Y_{n+1}$ can be expressed by \cref{eq:3e16}:
\begin{equation}
    Y_{n+1}=\sqrt{W_{n+1}^2(\alpha s_{n}^2+\sum_{i=1}^pa_iY_{n+1-i}^2)} \label{eq:3e16}
\end{equation}
\begin{remark}Even though we do not include the effect of $Y_t$ when we build $H_n$, the expression of $Y_{n+1}$ still contains the current value $Y_n$. It means the GE-NoVaS-without-$\beta$ method does not disobey the rule of causal prediction.
\end{remark}

Similarly, our proposed GA-NoVaS method can also be offered in a variant without $\beta$ term. \cref{4e9v,4e11} without $\beta$ term can be represented by following equations:
\begin{equation}
     W_t = \frac{Y_t}{\sqrt{ \alpha s_{t-1}^2 + \sum_{i = 1}^{q}\Tilde{c}_iY_{t-i}^2 }}~;~Y_t = \sqrt{W_t^2(\alpha s_{t-1}^2+\sum_{i=1}^q\Tilde{c}_iY_{t-i}^2)} \label{4e21}
\end{equation}
One thing should be mentioned here is that $\{\Tilde{c}_1,\cdots,\Tilde{c}_q\}$ represents $\{a_1,a_1b_1$ $,a_1b_1^{2},$ $\cdots,a_1b_1^{q-1} \}$ scaled by timing a scalar $\frac{1-\alpha}{\sum_{j=1}^{q}a_1b_1^{j-1}}$. Besides, $\alpha + \sum_{i=1}^{q}\Tilde{c}_i = 1$ is required to satisfy the variance-stabilizing requirement and the optimal combination of $a_1,b_1$ is selected by minimizing $\abs{KURT(W_t)-3}$ to satisfy the normalizing requirement. For GE-NoVaS- and GA-NoVaS-without-$\beta$ methods, we can still express $Y_{n+h}$ as a function of $\{W_{n+1},\cdots,W_{n+h}\}$ and repeat the aforementioned procedure to get $L_1$ and $L_2$ predictors. For example, we can derive the expression of $Y_{n+h}$ using the GA-NoVaS-without-$\beta$ method:

\begin{equation}
    Y_{n+h} = f_{\text{GA-without-$\beta$}}(W_{n+1},\cdots,W_{n+h};\mathscr{F}_{n})~;~\text{for any}~h\geq 1. \label{4e22}
\end{equation}

\begin{remark}[Slight computational efficiency of removing $\beta$]\label{remark3.2}
Note that the suggestion of removing $\beta$ can also lead a less time-complexity of the existing GE-NoVaS and newly proposed GA-NoVaS methods. The reason for this is simple: Recall $1/\sqrt{\beta}$ is required to be larger or equal to 3 for making $\{W_t\}$ have enough large range, i.e., $\beta$ is required to be less or equal to 0.111. However, the optimal combination of NoVaS coefficients may not render a suitable $\beta$. For this situation, we need to increase the time-series order ($p$ or $q$) and repeat the normalizing and variance-stabilizing process till $\beta$ in the optimal combination of coefficients is appropriate. This replication process definitely increases the computation workload.
\end{remark}

\subsection{Connection of two parsimonious methods}\label{ssec:connection}
In this subsection, we reveal that GE-NoVaS-without-$\beta$ and GA-NoVaS-without-$\beta$ methods actually have a same structure. The difference between these two methods lies in the region of free parameters. For observing this phenomenon, let us consider scaled coefficients of GA-NoVaS-without-$\beta$ method except $\alpha$:
\begin{equation}
\left\{ \frac{(1-\alpha)b_1^{i-1}}{\sum_{j=1}^{q}b_1^{j-1}}\right\}_{i=1}^{q} =\left\{ \frac{(1-\alpha)b_1^{i}}{\sum_{j=1}^{q}b_1^{j}}\right\}_{i=1}^{q} \label{eq:3.19}
\end{equation}
Recall parameters of GE-NoVaS-without-$\beta$ method except $\alpha$ implied by \cref{3.2e6} are:
\begin{equation}
    \left\{ \frac{(1-\alpha)e^{-ci}}{\sum_{j=1}^pe^{-cj}} \right\}_{i=1}^{p} \label{eq:3.20}
\end{equation}

Observing above two equations, although we can discover that \cref{eq:3.19} and \cref{eq:3.20} are equivalent if we set $b_1$ being equal to $e^{-c}$, these two methods are still slightly different since regions of $b_1$ and $c$ play a role in the process of optimization. The complete region of $c$ could be $(0,\infty)$. However, \cite{politis2015modelfreepredictionprinciple} pointed out that $c$ can not take a large value\footnote{When $c$ is large, $a_i \approx 0$ for all $i > 0$. It is hard to make the kurtosis of transformed series be 3.} and the region of $c$ should be an interval of the type $(0,m)$ for some $m$. In other words, a formidable search problem for finding the optimal $c$ is avoided by choosing such trimmed interval. On the other hand, $b_1$ is explicitly searched from $(0,1)$ which is corresponding with $c$ taking values from $(0,\infty)$. Likewise, applying the GA-NoVaS-without-$\beta$ method, the aforementioned burdensome search problem is also eliminated. Moreover, we can build a transformation based on the whole available region of unknown parameter. In spite of the fact that GE-NoVaS-without-$\beta$ and GA-NoVaS-without-$\beta$ methods have indistinguishable prediction performance for most of data analysis cases, we argue that the GA-NoVaS-without-$\beta$ method is more stable and reasonable than the GE-NoVaS-without-$\beta$ method since it is a more complete technique viewing the available region of parameter. Moreover, GA-NoVaS-without-$\beta$ method achieves significantly better prediction performance for some cases, see more details from \hyperref[appendix:a]{Appendix A}.

\subsection{Algorithms of new methods}\label{ssc:algorithm}
\noindent In \cref{ssc:garchnovas,subsec:parsimoniousvariant}, we exhibit the GA-NoVaS method and its parsimonious variant. In this section, we provide algorithms of these two methods. For the GA-NoVaS method, unknown parameters $\beta, a_1, b_1$ are selected from three grids of possible values to normalize $\{W_t;~t = q+1,\cdots,n\}$ in \cref{4e9}. If our goal is the $h$-step ahead prediction of $g(Y_{n+h})$ using past $\{Y_t;~t=1,\cdots,n\}$, the algorithm of the GA-NoVaS method can be summarized in \cref{algori1}.

\begin{algorithm}[htbp]
\caption{the $h$-step ahead prediction for the GA-NoVaS method}
\centering
\label{algori1}
  \centering
  \begin{tabular} {p{29pt}p{280pt}}   
    Step 1 & Define a grid of possible $\alpha$ values, $\{\alpha_k;~ k = 1,\cdots,K\}$, three grids of possible $\beta$, $a_1$, $b_1$ values. Fix $\alpha = \alpha_k$, then calculate the optimal combination of $\beta,a_1,b_1$ of the GA-NoVaS method.\\
    Step 2 & Derive the analytic form of \cref{4e13} using $\{\beta, a_1, b_1, \alpha_k\}$ from the first step.\\
    Step 3 & Generate $\{W_{n+1,m},\cdots, W_{n+h,m}\}_{m=1}^{M}$ from a trimmed standard normal distribution or empirical distribution $\hat{F}_w$. Plug $\{W_{n+1,m},\cdots, W_{n+h,m}\}_{m=1}^{M}$ into the analytic form of \cref{4e13} to obtain $M$ pseudo-values $\{Y_{n+h,m}\}_{m=1}^{M}$.\\
    Step 4 & Calculate the optimal predictor $g(\hat{Y}_{n+h})$ by taking the sample mean (under $L_2$ risk criterion) or sample median (under $L_1$ risk criterion) of the set $\{g(Y_{n+h,1}),\cdots,g(Y_{n+h,M})\}$.\\
    Step 5 & Repeat above steps with different $\alpha$ values from $\{\alpha_k;~ k = 1,\cdots,K\}$ to get $K$ prediction results.   \\ 
  \end{tabular}
\end{algorithm}
If we want to apply the GA-NoVaS-without-$\beta$ method, we just need to change \cref{algori1} a little bit. The difference between \cref{algori1,algori2} is the optimization of $\beta$ term being removed. The optimal combination of $a_1,b_1$ is still selected based on the normalizing and variance-stabilizing purpose. In our experiment setting, we choose regions of $\beta,a_1,b_1$ being $(0,1)$ and set a 0.02 grid interval to find all parameters. Besides, for the GA-NoVaS method, we also make sure that the sum of $\beta,a_1,b_1$ is less than 1 and the coefficient of $Y_t^{2}$ is the largest one. \\ 
\begin{algorithm}[H]
\centering
\caption{the $h$-step ahead prediction for the GA-NoVaS-without-$\beta$}
\label{algori2}
\hspace{0.5cm}
  \centering
  \begin{tabular} {p{29pt}p{280pt}}   
    \centering Step 1 & Define a grid of possible $\alpha$ values, $\{\alpha_k;~ k = 1,\cdots,K\}$, two grids of possible $a_1$, $b_1$ values. Fix $\alpha = \alpha_k$, then calculate the optimal combination of $a_1,b_1$ of the GA-NoVaS-without-$\beta$ method.\\
    \centering Steps 2-5  & Same as \cref{algori1}, but $\{W_{n+1,m},\cdots, W_{n+h,m}\}_{m=1}^{M}$ are plugged into the analytic form of \cref{4e22} and the standard normal distribution does not need to be truncated.
    \\
  \end{tabular}
\end{algorithm}

\section{Simulation}\label{sec:simu}

\noindent In simulation studies, for controlling the dependence of prediction performance on the length of the dataset, 16 datasets (2 from each settings) are generated from different GARCH(1,1)-type models separately and the size of each dataset is 250 (short data mimics 1-year of econometric data) or 500 (large data mimics 2-years of econometric data).
\\
\\
\textbf{Model 1:}  Time-varying GARCH(1,1) with Gaussian errors\\
$X_t = \sigma_t\epsilon_t,~\sigma_t^2 = \omega_{0,t} + \beta_{1,t}\sigma_{t-1}^2+\alpha_{1,t}X_{t-1}^2,~\{\epsilon_t\}\sim i.i.d.~N(0,1)$\\
$g_t = t/n; \omega_{0,t}= -4sin(0.5\pi g_t)+5; \alpha_{1,t} = -1(g_t-0.3)^2 + 0.5; \beta_{1,t} = 0.2sin(0.5\pi g_t)+0.2,~n = 250~\text{or}~500$\\
\textbf{Model 2:} Another time-varying GARCH(1,1) with Gaussian errors\\
$X_t = \sigma_t\epsilon_t,~\sigma_t^2 = 0.00001 + \beta_{1,t}\sigma_{t-1}^2+\alpha_{1,t}X_{t-1}^2,~\{\epsilon_t\}\sim i.i.d.~N(0,1)$\\
$g_t = t/n$; $\alpha_{1,t} = 0.1 - 0.05g_t$; $\beta_{1,t} = 0.73 + 0.2g_t,~n = 250~\text{or}~500$\\
\textbf{Model 3:} Standard GARCH(1,1) with Gaussian errors\\
$X_t = \sigma_t\epsilon_t,~\sigma_t^2 = 0.00001 + 0.73\sigma_{t-1}^2+0.1X_{t-1}^2,~\{\epsilon_t\}\sim i.i.d.~N(0,1)$\\
\textbf{Model 4:} Standard GARCH(1,1) with Gaussian errors\\
$X_t = \sigma_t\epsilon_t,~\sigma_t^2 = 0.00001 + 0.8895\sigma_{t-1}^2+0.1X_{t-1}^2,~\{\epsilon_t\}\sim i.i.d.~N(0,1)$\\
\textbf{Model 5:} Standard GARCH(1,1) with Student-$t$ errors\\
$X_t = \sigma_t\epsilon_t,$ $~\sigma_t^2 = 0.00001 + 0.73\sigma_{t-1}^2+0.1X_{t-1}^2,$\\ $~\{\epsilon_t\}\sim i.i.d.~t$ $\text{distribution with five degrees of freedom}$\\
\textbf{Model 6:} Exponential GARCH(1,1) with Gaussian errors\\
$X_t = \sigma_t\epsilon_t,~\log(\sigma_t^2) = 0.00001 + 0.8895\log(\sigma^2_{t-1})+0.1\epsilon_{t-1}+0.3(\abs{\epsilon_{t-1}}-E\abs{\epsilon_{t-1}}),$\\$~\{\epsilon_t\}\sim i.i.d.~N(0,1)$\\
\textbf{Model 7:} GJR-GARCH(1,1) with Gaussian errors\\
$X_t = \sigma_t\epsilon_t,~\sigma_t^2 = 0.00001 + 0.5\sigma^2_{t-1}+0.5X_{t-1}^2-0.5I_{t-1}X_{t-1}^2,~\{\epsilon_t\}\sim i.i.d.~N(0,1)\\
I_{t} = 1~\text{if}~ X_t \leq 0; I_{t} = 0~ \text{otherwise}$\\
\textbf{Model 8:} Another GJR-GARCH(1,1) with Gaussian errors\\
$X_t = \sigma_t\epsilon_t,~\sigma_t^2 = 0.00001 + 0.73\sigma^2_{t-1}+0.1X_{t-1}^2+0.3I_{t-1}X_{t-1}^2,~\{\epsilon_t\}\sim i.i.d.~N(0,1)\\
I_{t} = 1~\text{if}~ X_t \leq 0; I_{t} = 0~ \text{otherwise}$\\

\textit{Model description:} Models 1 and 2 present a time-varying GARCH model where coefficients $a_0, a_1, b_1$ change over time slowly. They differ significantly in the intercept term of $\sigma_t^2$ as we intentionally keep it low in the second setting. Models 3 and 4 are from a standard GARCH where in Model 4 we wanted to explore a scenario that $\alpha_1+\beta_1$ is very close to 1 and thus mimic what would happen for the iGARCH situation. Model 5 allows for the error distribution to come from a student-$t$ distribution instead of the Gaussian distribution. Note that, for a fair competition, we chose Models 2 to 5 same as simulation settings of \citep{chen2019optimal}. Models 6, 7 and 8 present different types of GARCH models. These settings allow us to check robustness of our methods against model misspecification. In a real world, it is hard to convincingly say if the data obeys one particular type of GARCH model, so we want to pursue this exercise to see if our methods are satisfactory no matter what the underlying distribution and the GARCH-type model are. This approach to test the performance of a method under model misspecification is quite standard, see \cite{olubusoye2016misspecification} used data generated from a specifically true model to estimate other GARCH-type models and test the forecasting performance, and \cite{bellini2008misspecification} investigated the impact of misspecification of innovations in fitting GARCH models.

\textit{Window size:} Using these datasets, we perform 1-step, 5-steps and 30-steps ahead time-aggregated POOS predictions. For measuring different methods' prediction performance on larger datasets (i.e., data size is 500), we use 250 data as a window to do predictions and roll this window through the whole dataset. For evaluating different methods' performance on smaller datasets (i.e., data size is 250), we use 100 data as a window. 

Note that log-returns can be calculated from equation shown below:
\begin{equation}
    Y_t = 100\times \log(X_{t+1}/X_t) ~;~\text{for}~ t = 1,\cdots,499~\text{or}~t = 1,\cdots,249. \label{Eq:4.1}
\end{equation}
Where, $\{X_t\}_{t = 1}^{250}$ and $\{X_t\}_{t = 1}^{500}$ are 1-year and 2-years price series, respectively. Next, we can define time-aggregated predictions of squared log-returns as:
\begin{equation}
\begin{split}
    \bar{Y}_{k,1}^2 = \hat{Y}_{k+1}^2,~k=250,\cdots,498 ~\text{or}~k=100,\cdots,248\\
    \bar{Y}_{i,5}^2 = \frac{1}{5}\sum_{m=1}^5\hat{Y}^2_{i+m},~i = 250,\cdots,494~\text{or}~i=100,\cdots,244\\
    \bar{Y}_{j,30}^2 = \frac{1}{30}\sum_{m=1}^{30}\hat{Y}^2_{j+m},~j = 250,\cdots,469~\text{or}~j=100,\cdots,219 \label{4e17}
\end{split}
\end{equation}
In \cref{4e17}, $\hat{Y}_{k+1}^2,\hat{Y}_{i+m}^2,\hat{Y}_{j+m}^2$ are single point predictions of realized squared log-returns by NoVaS-type methods or benchmark method; $\bar{Y}_{k,1}^2$, $\bar{Y}_{i,5}^2$ and $\bar{Y}_{j,30}^2$ represent 1-step, 5-steps and 30-steps ahead aggregated predictions, respectively. More specifically, for exploring the performance of three different prediction lengths with large data size, we roll the 250 data points window through the whole dataset, i.e., use $\{Y_1,\cdots,Y_{250}\}$ to predict $Y_{251}^2,\{Y_{251}^2,\cdots,Y_{255}^2\}$ and $\{Y_{251}^2,\cdots,Y_{280}^2\}$; then use $\{Y_2,\cdots,Y_{251}\}$ to predict $Y_{252}^2,\{Y_{252}^2,\cdots,Y_{256}^2\}$ and $\{Y_{252}^2,\cdots,Y_{281}^2\}$, for 1-step, 5-steps and 30-steps aggregated predictions respectively, and so on. For exploring the performance of three different prediction lengths with small data size, we roll the 100 data points window through the whole dataset, i.e., use $\{Y_1,\cdots,Y_{100}\}$ to predict $Y_{101}^2,\{Y_{101}^2,\cdots,Y_{105}^2\}$ and $\{Y_{101}^2,\cdots,Y_{130}^2\}$; then use $\{Y_2,\cdots,Y_{101}\}$ to predict $Y_{102}^2,\{Y_{102}^2,\cdots,Y_{106}^2\}$ and $\{Y_{102}^2,\cdots,Y_{131}^2\}$, for 1-step, 5-steps and 30-steps aggregated predictions respectively, and so on. For example, with window size being 30, we perform time-aggregated predictions on a large dataset 220 times. Taking this strategy, we can exhaust the information contained in the dataset and investigate the forecasting performance continuously.  

To measure different methods' forecasting performance, we compare predictions with realized values based on \cref{eq:4.1}. 
\begin{equation}
    P = \sum_{l}(\bar{Y}_{l,h}^2-\sum_{m=1}^h(Y_{l+m}^2/h))^2~;~l \in \{k,i,j\}\label{eq:4.1}
\end{equation}
In \cref{eq:4.1}, setting $l = k,i,j$ means we consider 1-step, 5-steps and 30-steps ahead time-aggregated predictions respectively; $\bar{Y}_{l,h}^2$ is the $h$-step ($h\in\{1,5,30\}$) ahead time-aggregated volatility prediction, defined in \cref{4e17}; $\sum_{m=1}^h(Y_{l+m}^2/h)$ is the corresponding true aggregated value calculated from realized squared log-returns. For comparing various Model-free methods with the traditional method, we set the benchmark method as fitting one GARCH(1,1) model directly (GARCH-direct).

\textit{Different variants of methods:} Note that we can perform GE-NoVaS-type and GA-NoVaS-type methods to predict $Y_{n+h}$ by generating $\{W_{n+1,m},\cdots, W_{n+h,m}\}_{m=1}^{M}$ from a standard normal distribution or the empirical distribution of $\{W_t\}$ series, then we can calculate the optimal predictor based on $L_1$ or $L_2$ risk criterion. It means each NoVaS-type method possesses four variants.  

When we perform POOS forecasting, we do not know which $\alpha$ is optimal. Thus, we perform every NoVaS variants using $\alpha$ from eight potential values $\{0.1, 0.2, \cdots,0.8\}$ and then pick the optimal result. For simplifying the presentation, we further select the final prediction from optimal results of four variants of a NoVaS method and use this result to be the best prediction to which each NoVaS method can reach. Applying this procedure means we take a computationally heavy approach to compare different methods' potentially best performance. However, it also means we want to challenge newly proposed methods at a maximum level, so as to see if they can beat even the best-performing scenario of the current GE-NoVaS method. 

\subsection{Simulation results}\label{ssec:simuresults}
\noindent In this subsection, we compare the performance of our new methods (GA-NoVaS and GA-NoVaS-without-$\beta$) with GARCH-direct and existing GE-NoVaS methods on forecasting 250 and 500 simulated data. Results are tabulated in \cref{5t1}.

\subsubsection{Simulation results of Models 1 to 5}\label{sssec:simuresultsmoeld1-5}
\noindent From \cref{5t1}, we clearly find NoVaS-type methods outperform the GARCH-direct method. Especially for using the 500 Model-1 data to do 30-steps ahead aggregated prediction, the performance of the GARCH-direct method is terrible. NoVaS-type methods are almost 30 times better than the GARCH-direct method. This means that the normal prediction method may be spoiled by error accumulation problem when long-term predictions are required. On the other hand, Model-free methods can avoid this problem.

In addition to the overall advantage of NoVaS-type methods over GARCH-direct method, we find the GA-NoVaS method is generally better than the GE-NoVaS method for both short and large data. This conclusion is two-fold: (1) The time of the GA-NoVaS being the best method is more than the GE-NovaS method; (2) Since we want to compare the forecasting ability of GE-NoVaS and GA-NoVaS methods, we use $*$ symbol to represent cases where the GA-NoVaS method works at least 10$\%$ better than the GE-NoVaS method or inversely the GE-NoVaS method is 10$\%$ better. We can find there is no case to support that the GE-NoVaS works better than GA-NoVaS with as least 10$\%$ improvement. On the other hand, the GA-NoVaS method achieves significant improvement when long-term predictions are required. Moreover, the GA-NoVaS-without-$\beta$ dominates other two NoVaS-type methods.

\subsubsection{Models 6 to 8: Different GARCH specifications}\label{sssc:simudifferent}
\noindent Since the main crux of Model-free methods is how such non-parametric methods are robust to underlying data-generation processes, here we explore other GARCH-type data generations. The GA-NoVaS method is based upon GARCH model, so it is interesting to explore whether even these methods can sustain a different type of true underlying generation and can in general outperform existing methods. Results for Models 6 to 8 are tabulated in \cref{5t1}.

In general, NoVaS-type methods still outperform the GARCH-direct method for these cases. Although the forecasting ability of GE-NoVaS and GA-NoVaS for large data is indistinguishable, the GA-NoVaS is obviously better for taking short data size. For example, the GA-NoVaS method brings around 20$\%$ improvement compared with the GE-NoVaS method for 30-steps ahead aggregated prediction of 250 Model-6 simulated data. Doing better prediction with past data that is shorter in size is always a significant challenge and thus it is valuable to discover the GA-NoVaS method has superior performance for this scenario. Not surprisingly, the GA-NoVaS-without-$\beta$ method still keeps great performance.

\subsection{Simulation summary}\label{ssc:simusmall}
\noindent Through deploying simulation data analysis, we find GA-NoVaS-type methods can sustain great performance against short data and model misspecification. Overall, our new methods outperform the GE-NoVaS method and can render notable improvement for some cases when long-term predictions are desired. 

\begin{table}[htbp]
\caption{Comparison results of using 500 and 250 simulated data}
\label{5t1}
\begin{adjustbox}{width=1\textwidth}
\small
\begin{tabular}{lcccclcccc}
  \toprule
 \textbf{500} size & \thead{\small GE}  & \thead{\small GA} &  \thead{\small P-GA}  & \thead{\small GARCH}  & \textbf{250} size & \thead{\small GE}  & \thead{\small GA} &  \thead{\small P-GA}  & \thead{\small GARCH}\\ 
  \midrule
% Table generated by Excel2LaTeX from sheet 'Sheet2'

    M1-1step  & 0.89258 & 0.88735 &  \textbf{0.84138} & 1.00000 &  M1-1step  & 0.91538 & 0.9112 &  \textbf{0.83034} & 1.00000 \\ [2pt]
        M1-5steps  & 0.40603 & 0.40296 &  \textbf{0.40137} & 1.00000 &     M1-5steps  & 0.49169 & 0.48479 &  \textbf{0.43247} & 1.00000 \\[2pt]
        M1-30steps  & 0.03368 & 0.03294 &  \textbf{0.03290} & 1.00000 &     M1-30steps  & 0.25009 & 0.24752 &  \textbf{0.23035} & 1.00000 \\[2pt]
        M2-1step  &  \textbf{0.95689} & 0.96069 & 0.99658 & 1.00000 &     M2-1step  & 0.91369 & 0.91574 &  \textbf{0.87614} & 1.00000 \\[2pt]
        M2-5steps  & 0.89981 &  \textbf{0.89739} & 0.9198 & 1.00000 &     M2-5steps  & 0.61001 & 0.61094 &  \textbf{0.51712} & 1.00000 \\[2pt]
        M2-30steps  & 0.63126 & 0.64042 &  \textbf{0.48396} & 1.00000 &     M2-30steps  & 0.7725 &  \textbf{0.74083} & 0.75251 & 1.00000 \\[2pt]
        M3-1step  & 0.99938 & 1.00150 &  \textbf{0.98407} & 1.00000 &     M3-1step  & 0.97796 & 0.96632 &  \textbf{0.93693} & 1.00000 \\[2pt]
        M3-5steps  & 0.98206 & 0.96088 &  \textbf{0.94073} & 1.00000 &     M3-5steps  & 0.98127 &  \textbf{0.97897} & 0.99977 & 1.00000 \\[2pt]
        M3-30steps  & 1.10509 & 1.03683 &  \textbf{0.90855} & 1.00000 &     M3-30steps  & 1.38353 &  \textbf{0.89001*} & 0.99818 & 1.00000 \\[2pt]
        M4-1step  & 0.98713 &  \textbf{0.98466} & 0.9964 & 1.00000 &     M4-1step  & 0.99183 & 0.95698 &  \textbf{0.92811} & 1.00000 \\[2pt]
        M4-5steps  & 0.95382 & 0.95362 &  \textbf{0.95338} & 1.00000 &     M4-5steps  & 0.77088 & 0.72882 &  \textbf{0.67894} & 1.00000 \\[2pt]
        M4-30steps  & 0.75811 & 0.69208 &  \textbf{0.67594} & 1.00000 &     M4-30steps  & 0.79672 &  \textbf{0.6095*} & 0.81115 & 1.00000 \\[2pt]
        M5-1step  & 0.96940 &  \textbf{0.94066} & 0.97151 & 1.00000 &     M5-1step  & 0.83631 & 0.84134 & 0. \textbf{79075} & 1.00000 \\[2pt]
        M5-5steps  & 0.84751 &  \textbf{0.72806*} & 0.82747 & 1.00000 &     M5-5steps  & 0.38296 & 0.38034 &  \textbf{0.35155} & 1.00000 \\[2pt]
        M5-30steps  & 0.49669  &\textbf{ 0.24318*} & 0.47311 & 1.00000 &     M5-30steps  & 0.00199 & 0.002 &  \textbf{0.00194} & 1.00000 \\[2pt]
        M6-1step  & 1.00175 & 1.00514 &  \textbf{0.93509} & 1.00000 &     M6-1step  & 0.95939 & 0.96499 &  \textbf{0.93863} & 1.00000 \\[2pt]
        M6-5steps  & 0.93796 & 0.94249 & \textbf{ 0.80311} & 1.00000 &     M6-5steps  & 0.93594 & 0.97101 &  \textbf{0.85851} & 1.00000 \\[2pt]
        M6-30steps  & 0.50740 & 0.51350 &  \textbf{0.41112} & 1.00000 &     M6-30steps  & 0.84401 &  \textbf{0.67272*} & 0.7042 & 1.00000 \\[2pt]
        M7-1step  & 0.98857 & 0.98737 &  \textbf{0.95932} & 1.00000 &     M7-1step  & 0.84813 & 0.83628 &  \textbf{0.83216} & 1.00000 \\[2pt]
        M7-5steps  & 0.85539 & 0.85371 &  \textbf{0.85127} & 1.00000 &     M7-5steps  & 0.50849 & 0.50126 &  \textbf{0.4802} & 1.00000 \\[2pt]
        M7-30steps  &  \textbf{0.68202} & 0.68314 & 0.71391 & 1.00000 &     M7-30steps  & 0.06832 & 0.06817 &  \textbf{0.06507} & 1.00000 \\[2pt]
        M8-1step  & 0.96001 & 0.96463 &  \textbf{0.93452} & 1.00000 &     M8-1step  &  \textbf{0.79561} & 0.79994 & 0.8334 & 1.00000 \\[2pt]
        M8-5steps  & 0.97019 & 0.98184 &  \textbf{0.93178} & 1.00000 &     M8-5steps  & 0.48028 & 0.47244 &  \textbf{0.45665} & 1.00000 \\[2pt]
        M8-30steps  &  \textbf{0.30593} & 0.31813 & 0.33853 & 1.00000 &     M8-30steps  & 0.00977 &  \textbf{0.00942} & 0.00983 & 1.00000 \\[2pt]
     \bottomrule
\end{tabular}
\end{adjustbox}
\\
\tiny \textit{Note:} Column names ``GA'' and ``GE'' represent GE-NoVaS and GA-NoVaS methods, respectively; ``GARCH'' means GARCH-direct method; ``P-GA'' means GA-NoVaS-without-$\beta$ method. The benchmark is the GARCH-direct method, so numerical values in the table corresponding to GARCH-direct method are 1. Other numerical values are relative values compared to the GARCH-direct method. ``$Mi\text{-}j$''steps denotes using data generated from the Model $i$ to do $j$ steps ahead time-aggregated predictions. The bold value means that the corresponding method is the optimal choice for this data case. Cell with $*$ means the GA-NoVaS method is at least 10$\%$ better than the GE-NoVaS method or inversely the GE-NoVaS method is at least 10$\%$ better.
\end{table}

\section{Real-world data analysis}\label{sec:real data}
\noindent From \cref{sec:simu}, we have found that NoVaS-type methods have great performance on dealing with different simulated datasets. However, no methodological proposal is complete unless one verifies it on several real-world datasets. This section is devoted to explore, in the context of real datasets forecasting, whether NoVaS-type methods can provide good long-term time-aggregated forecasting ability and how our new methods are compared to the existing Model-free method.

For performing an extensive analysis and subsequently acquiring a convincing conclusion, we use three types of data--stock, index and currency data--to do predictions. Moreover, as done in simulation studies, we apply this exercise on two different lengths of data. For building large datasets (2-years period data), we take new data which come from Jan.2018 to Dec.2019 and old data which come from around 20 years ago, separately. The dynamics of these econometric datasets have changed a lot in the past 20 years and thus we wanted to explore whether our methods are good enough for both old and new data. Subsequently, we challenge our methods using short (1-year period) real-life data. Finally, we also do forecasting using volatile data, i.e., data from Nov. 2019 to Oct. 2020. Note that economies across the world went through a recession due to the COVID-19 pandemic and then slowly recovered during this time-period, typically these sort of situations introduce systematic perturbation in the dynamics of econometric datasets. We wanted to see if our methods can sustain such perturbations or abrupt changes. 

\subsection{Old and new 2-years data}\label{ssec:realdatanormalperiod2years}
\noindent For mimicking the 2-years period data, we adopt several stock datasets with 500 data size to do forecasting. In summary, we still compare different methods' performance on 1-step, 5-steps and 30-steps ahead POOS time-aggregated predictions. Performing the similar procedure as which we did in \cref{sec:simu}, all results are shown in \cref{6t1}. We can clearly find NoVaS-type methods still outperform the GARCH-direct method. Additionally, although the GE-NoVaS method is indistinguishable with the  GA-NoVaS method, our new method is more robust than the GE-NoVaS method, see the 30-steps ahead prediction of old two-years BAC and MSFT cases. We can also notice that the GA-NoVaS-without-$\beta$ method is more robust than other two NoVaS methods. The $\beta$-removing idea proposed by \cite{wu2021boosting} is substantiated again. 

Since the main goal of this article is offering a new type of NoVaS method which has better performance than the GE-NoVaS method for dealing with short and volatile data, we provide more extensive data analysis to support our new methods in next sections.  
\begin{table}[htbp]
  \caption{Comparison results of using old and new 2-years data}
  \label{6t1}
\begin{adjustbox}{width=1\textwidth}
\centering
\small
\begin{tabular}{lcccclcccc}
  \toprule
 \thead{Old \\2-years} & \thead{\small GE}  & \thead{\small GA} & \thead{\small P-GA}   & \thead{\small GARCH} & \thead{New \\2-years}& \thead{\small GE}  & \thead{\small GA} & \thead{\small P-GA}   & \thead{\small GARCH} \\ 
  \midrule

    AAPL-1step  & 0.99795 & 0.99236 & \textbf{0.97836} & 1.00000 & AAPL-1step  & 0.80150 & \textbf{0.79899} & 0.79915 & 1.00000 \\[2pt]
        AAPL-5steps  & 1.04919 & 1.04800 & \textbf{0.96999} & 1.00000 &     AAPL-5steps  & 0.41405 & 0.42338 & \textbf{0.40427} & 1.00000 \\[2pt]
        AAPL-30steps  & 1.12563 & 1.21986 & \textbf{0.96174} & 1.00000 &     AAPL-30steps  & \textbf{0.13207} & 0.14046 & 0.14543 & 1.00000 \\[2pt]
        BAC-1step  & \textbf{0.99889} & 1.00396 & 1.02780 & 1.00000 &     BAC-1step  & 0.98393 & 0.99164 & \textbf{0.96542} & 1.00000 \\[2pt]
        BAC-5steps  & 1.04424 & 1.02185 & \textbf{0.99399} & 1.00000 &     BAC-5steps  & 0.98885 & 1.01480 & \textbf{0.91857} & 1.00000 \\[2pt]
        BAC-30steps  & 1.32452 & 1.13887\textbf{*} & 1.00363 & \textbf{1.00000} &     BAC-30steps  & 1.14111 & 1.03657 & \textbf{0.88596} & 1.00000 \\[2pt]
        MSFT-1step  & 0.98785 & 0.98598 & \textbf{0.96185} & 1.00000 &     MSFT-1step  & 0.98405 & 0.98630 & \textbf{0.96374} & 1.00000 \\[2pt]
        MSFT-5steps  & 1.00236 & 1.00096 & \textbf{0.95271} & 1.00000 &     MSFT-5steps  & 0.65027 & 0.67005 & \textbf{0.64278} & 1.00000 \\[2pt]
        MSFT-30steps  & 1.25272 & 1.09881\textbf{*} & \textbf{0.88515} & 1.00000 &     MSFT-30steps  & \textbf{0.19767} & 0.20060 & 0.21473 & 1.00000 \\[2pt]
        MCD-1step  & 1.01845 & 1.00789 & \textbf{0.99005} & 1.00000 &     MCD-1step  & 0.99631 & 0.99539 & \textbf{0.98035} & 1.00000 \\[2pt]
        MCD-5steps  & 1.11249 & 1.07748 & \textbf{0.97777} & 1.00000 &     MCD-5steps  & 0.95403 & 0.95327 & \textbf{0.91317} & 1.00000 \\[2pt]
        MCD-30steps  & 1.76385 & 1.69757 & \textbf{0.99418} & 1.00000 &     MCD-30steps  & 0.75730 & 0.75361 & \textbf{0.74557} & 1.00000 \\[2pt]
    \bottomrule
    \end{tabular}
   \end{adjustbox}
    \\
    
    \tiny \textit{Note:} Column names ``GA'' and ``GE'' represent GE-NoVaS and GA-NoVaS methods, respectively; ``GARCH'' means GARCH-direct method; ``P-GA'' means GA-NoVaS-without-$\beta$ method. The benchmark is the GARCH-direct method, so numerical values in the table corresponding to GARCH-direct method are 1. Other numerical values are relative values compared to the GARCH-direct method. The bold value means that the corresponding method is the optimal choice for this data case. Cell with $*$ means the GA-NoVaS method is at least 10$\%$ better than the GE-NoVaS method or inversely the GE-NoVaS method is at least 10$\%$ better.
\end{table}

\subsection{2018 and 2019 1-year data }\label{ssec:realdatanormalperiod1year}
\noindent For challenging our new methods in contrast to other methods for small real-life datasets, we separate every new 2-years period data in \cref{ssec:realdatanormalperiod2years} to two 1-year period datasets, i.e., separate four new stock datasets to eight samples. We believe evaluating the prediction performance using shorter data is a more important problem and thus we wanted to make our analysis very comprehensive. Therefore, for this exercise, we add 7 index datasets: Nasdaq, NYSE, Small Cap, Dow Jones, S$\&$P 500 , BSE and BIST; and two stock datasets: Tesla and Bitcoin into our analysis. 

From \cref{6t2} which presents prediction results of different methods on 2018 and 2019 stock data, we still observe that NoVaS-type methods outperform GARCH-direct method for almost all cases. Among different NoVaS methods, it is clear that our new methods are superior than the existing GE-NoVaS method. For 30-steps ahead predictions of 2018-BAC data, 2019-MCD and Tesla data, etc, the existing NoVaS method is even worse than the GARCH-direct method. On the other hand, the GA-NoVaS method is more stable than the GE-NoVaS method, e.g., 30$\%$ improvement is created for the 30-steps ahead prediction of 2018-BAC data. After applying the $\beta$-removing idea, the GA-NoVaS-without-$\beta$ significantly beats other methods for almost all cases.

From \cref{6t3} which presents prediction results of different methods on 2018 and 2019 index data, we can get the exactly same conclusion as before. NoVaS-type methods are far superior than the GARCH-direct and our new NoVaS methods outperform the existing GE-NoVaS method. Interestingly, the GE-NoVaS method is again beaten by the GARCH-direct method in some cases, such as 2019-Nasdaq, Smallcap and BIST. On the other hand, new methods still show more stable performance. Compared to the existing GE-NoVaS method, the GA-NoVaS-without-$\beta$ method creates around 60$\%$ improvement from the GE-NoVaS method on the 30-steps ahead prediction of 2019-BIST data. In addition, the GA-NoVaS method shows more than 10$\%$ improvement for all 2018-BSE cases.

Combining results presented in \cref{6t1,6t2,6t3}, our new methods present better performance than existing GE-NoVaS and GARCH-direct methods on dealing with small and large real-life data. The improvement generated by new methods using shorter sample size (1-year data) is more significant than using larger sample size (2-years data).

\begin{table}[H]
  \caption{Comparison results of using 2018 and 2019 stock data}
  \label{6t2}
    \begin{adjustbox}{width=1\textwidth}
\small
\begin{tabular}{lcccclcccc}
  \toprule 
 2018& \thead{GE} &   \thead{GA} &  \thead{P-GA} & \thead{GARCH} & 2019 & \thead{GE}   & \thead{GA} &  \thead{P-GA} &\thead{GARCH} \\
  \midrule
% Table generated by Excel2LaTeX from sheet 'Sheet2'

    MCD-1step  & 0.98514 & 0.97887 & \textbf{0.94412} & 1.00000 &     MCD-1step  & 0.95959 & 0.96348 & \textbf{0.94559} & 1.00000 \\[2pt]
        MCD-5steps  & 1.0272 & 1.02519 & \textbf{0.88151} & 1.00000 &     MCD-5steps  & 1.00723 & 1.01169 & \textbf{0.90602} & 1.00000 \\[2pt]
        MCD-30steps  & 0.62614 & 0.63992 & \textbf{0.61153} & 1.00000 &     MCD-30steps  & 1.05239 & 0.95714 & \textbf{0.77976} & 1.00000 \\[2pt]
        AAPL-1step  & 0.92014 & 0.92317 & \textbf{0.89283} & 1.00000 &     AAPL-1step  & 0.84533 & \textbf{0.81326} & 0.81872 & 1.00000 \\[2pt]
        AAPL-5steps  & 0.84798 & 0.73461\textbf{*} & \textbf{0.71233} & 1.00000 &     AAPL-5steps  & 0.85401 & 0.79254 & \textbf{0.68792} & 1.00000 \\[2pt]
        AAPL-30steps  & 0.38612 & \textbf{0.36324} & 0.37081 & 1.00000 &     AAPL-30steps  & 0.99043 & 0.99286 & \textbf{0.72892} & 1.00000 \\[2pt]
        BAC-1step  & 0.94952 & 0.93842 & 0\textbf{.92619} & 1.00000 &     BAC-1step  & 1.04272 & 1.04722 & \textbf{0.98605} & 1.00000 \\[2pt]
        BAC-5steps  & 0.83395 & 0.79158 & \textbf{0.72512} & 1.00000 &     BAC-5steps  & 1.22761 & 1.20195 & \textbf{0.95436} & 1.00000 \\[2pt]
        BAC-30steps  & 1.34367 & 0.90675\textbf{*} & \textbf{0.8763} & 1.00000 &     BAC-30steps  & 1.4502 & 1.41788 & 1.03482 & \textbf{1.00000} \\[2pt]
        MSFT-1step  & 0.91705 & \textbf{0.90936} & 0.95921 & 1.00000 &     MSFT-1step  & 1.03308 & 1.00101 & \textbf{0.95347} & 1.00000 \\[2pt]
        MSFT-5steps  & 0.74553 & 0.74267 & \textbf{0.74237} & 1.00000 &     MSFT-5steps  & 1.2234 & 1.18205 & \textbf{0.95417} & 1.00000 \\[2pt]
        MSFT-30steps  & 0.6699 & 0.6477 & \textbf{0.64717} & 1.00000 &     MSFT-30steps  & 1.2302 & 1.21337 & \textbf{0.98476} & 1.00000 \\[2pt]
        Tesla-1step  & 1.00181 & 0.96074 & \textbf{0.86238} & 1.00000 &     Tesla-1step  & 1.00428 & 1.01934 & \textbf{0.98955} & 1.00000 \\[2pt]
        Tesla-5steps  & 1.20383 & 1.13335 & 1.0156 & \textbf{1.00000} &     Tesla-5steps  & 1.0661 & 1.07506 & \textbf{0.96107} & 1.00000 \\[2pt]
        Tesla-30steps  & 1.97328 & 1.84871 & 1.25005 & \textbf{1.00000} &     Tesla-30steps  & 2.00623 & 1.71782\textbf{*} & \textbf{0.84366} & 1.00000 \\[2pt]
        Bitcoin-1step  & 0.99636 & 1.01731 & \textbf{0.97734} & 1.00000 &     Bitcoin-1step  & 0.89929 & 0.88914 & \textbf{0.87256} & 1.00000 \\[2pt]
        Bitcoin-5steps  & 1.02021 & 1.1188 & \textbf{0.93826} & 1.00000 &     Bitcoin-5steps  & 0.62312 & 0.63075 & \textbf{0.56789} & 1.00000 \\[2pt]
        Bitcoin-30steps  & \textbf{0.86649} & 0.95506 & 0.91364 & 1.00000 &     Bitcoin-30steps  & 0.00733 & 0.00749 & \textbf{0.00631} & 1.00000 \\[2pt]

       \bottomrule
    \end{tabular}
    \end{adjustbox}
        \\
    \tiny \textit{Note:} Column names ``GA'' and ``GE'' represent GE-NoVaS and GA-NoVaS methods, respectively; ``GARCH'' means GARCH-direct method; ``P-GA'' means GA-NoVaS-without-$\beta$ method. The benchmark is the GARCH-direct method, so numerical values in the table corresponding to GARCH-direct method are 1. Other numerical values are relative values compared to the GARCH-direct method. The bold value means that the corresponding method is the optimal choice for this data case. Cell with $*$ means the GA-NoVaS method is at least 10$\%$ better than the GE-NoVaS method or inversely the GE-NoVaS method is at least 10$\%$ better.
\end{table}

\begin{table}[H]
  \caption{Comparison results of using 2018 and 2019 index data}
   \setlength{\abovecaptionskip}{0pt}
  \label{6t3}
    \begin{adjustbox}{width=1\textwidth}
\small
\begin{tabular}{lcccclcccc}
  \toprule 
 2018 & \thead{GE} & \thead{GA} & \thead{P-GA}  & \thead{GARCH} & 2019& \thead{GE} & \thead{GA} & \thead{P-GA} & \thead{GARCH} \\ 
  \midrule
    Nasdaq-1step  & \textbf{0.91309} & 0.92303 & 0.92421 & 1.00000 &     Nasdaq-1step  & 0.99960 & 0.98950 & \textbf{0.93843} & 1.00000 \\[2pt]
        Nasdaq-5steps  & \textbf{0.76419} & 0.79718 & 0.78823 & 1.00000 &     Nasdaq-5steps  & 1.15282 & 1.09176 & \textbf{0.84051} & 1.00000 \\[2pt]
        Nasdaq-30steps  & 0.66520 & \textbf{0.65489} & 0.67389 & 1.00000 &     Nasdaq-30steps  & 0.68994 & 0.69846 & \textbf{0.59218} & 1.00000 \\[2pt]
        NYSE-1step  & 0.93509 & \textbf{0.93401} & 0.96619 & 1.00000 &     NYSE-1step  & 0.92486 & \textbf{0.91118} & 0.92193 & 1.00000 \\[2pt]
        NYSE-5steps  & 0.83725 & 0.79330 & \textbf{0.75822} & 1.00000 &     NYSE-5steps  & 0.86249 & 0.82114 & \textbf{0.71038} & 1.00000 \\[2pt]
        NYSE-30steps  & 0.75053 & \textbf{0.61443*} & 0.61830 & 1.00000 &     NYSE-30steps  & 0.22122 & 0.22173 & \textbf{0.18116} & 1.00000 \\[2pt]
        Smallcap-1step  & \textbf{0.90546} & 0.91346 & 0.91101 & 1.00000 &     Smallcap-1step  & 1.02041 & 1.00626 & \textbf{0.98482} & 1.00000 \\[2pt]
        Smallcap-5steps  & \textbf{0.72627} & 0.73955 & 0.73223 & 1.00000 &     Smallcap-5steps  & 1.15868 & 1.08929 & \textbf{0.85490} & 1.00000 \\[2pt]
        Samllcap-30steps  & 0.50005 & 0.46482 & \textbf{0.46312} & 1.00000 &     Samllcap-30steps  & 1.30467 & 1.28949 & \textbf{0.90360} & 1.00000 \\[2pt]
        Djones-1step  & 0.90932 & \textbf{0.90707} & 0.91192 & 1.00000 &     Djones-1step  & 0.96752 & \textbf{0.96433} & 0.96977 & 1.00000 \\[2pt]
        Djones-5steps  & 0.82480 & 0.79965 & \textbf{0.76226} & 1.00000 &     Djones-5steps  & 0.98725 & 0.93315 & \textbf{0.91238} & 1.00000 \\[2pt]
        Djones-30steps  & 0.72547 & \textbf{0.53021*} & 0.56854 & 1.00000 &     Djones-30steps  & 0.86333 & 0.85006 & \textbf{0.81803} & 1.00000 \\[2pt]
        SP500-1step  & 0.91860 & 0.91256 & \textbf{0.88405} & 1.00000 &     SP500-1step  & 0.96978 & 0.96526 & \textbf{0.93162} & 1.00000 \\[2pt]
        SP500-5steps  & 0.85108 & 0.77305 & \textbf{0.75646} & 1.00000 &     SP500-5steps  & 0.96704 & 0.94028 & \textbf{0.77434} & 1.00000 \\[2pt]
        SP500-30steps  & 0.88917 & \textbf{0.68156*} & 0.72104 & 1.00000 &     SP500-30steps  & 0.34389 & 0.34537 & \textbf{0.30127} & 1.00000 \\[2pt]
        BSE-1step  & 0.99942 & \textbf{0.88322*} & 0.92568 & 1.00000 &     BSE-1step  & 0.70667 & 0.70194 & \textbf{0.66667} & 1.00000 \\[2pt]
        BSE-5steps  & 0.92061 & \textbf{0.78484*} & 0.84408 & 1.00000 &     BSE-5steps  & 0.25675 & 0.25897 & \textbf{0.23603} & 1.00000 \\[2pt]
        BSE-30steps  & 0.52431 & \textbf{0.41010*} & 0.44092 & 1.00000 &     BSE-30steps  & 0.03764 & 0.03951 & \textbf{0.02888} & 1.00000 \\[2pt]
        BIST-1step  & 0.93221 & \textbf{0.92215} & 0.94138 & 1.00000 &      BIST-1step  & \textbf{0.96807} & 0.97209 & 0.98234 & 1.00000 \\[2pt]
        BIST-5steps  & 0.82149 & \textbf{0.79664} & 0.81417 & 1.00000 &      BIST-5steps  & 0.98944 & 1.03903 & \textbf{0.85370} & 1.00000 \\[2pt]
        BIST-30steps  & 1.34581 & 1.42233 & 1.09900 & \textbf{1.00000} &      BIST-30steps  & 2.21996 & 2.10562 & \textbf{0.85743} & 1.00000 \\[2pt]

       \bottomrule
    \end{tabular}
    \end{adjustbox}
        \\
    \tiny \textit{Note:} Column names ``GA'' and ``GE'' represent GE-NoVaS and GA-NoVaS methods, respectively; Column name ``GARCH'' means GARCH-direct method; ``P-GA'' means GA-NoVaS-without-$\beta$ method. The benchmark is the GARCH-direct method, so numerical values in the table corresponding to GARCH-direct method are 1. Other numerical values are relative values compared to the GARCH-direct method. The bold value means that the corresponding method is the optimal choice for this data case. Cell with $*$ means the GA-NoVaS method is at least 10$\%$ better than the GE-NoVaS method or inversely the GE-NoVaS method is at least 10$\%$ better.
\end{table}

\subsection{Volatile 1-year data}\label{ssec: realdatavolatileperiod}
\begin{figure}[htbp]
\centering
\includegraphics[width=9cm,height=6cm]{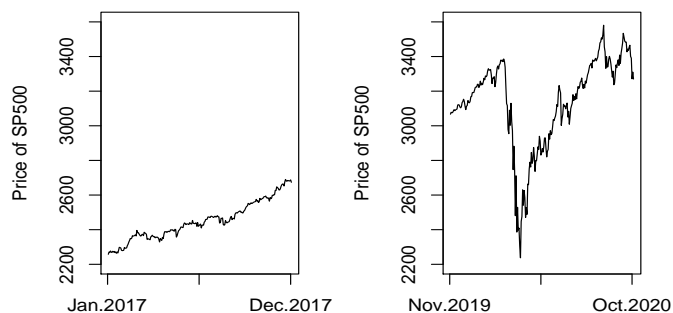}
\caption{The left subfigure depicts the price of SP500 from Jan.2017 to Dec.2017 which presents a slow growth; The right subfigure depicts the price of SP500 from Nov.2019 to Oct.2020}
\label{6f1}
\end{figure}

\noindent In this subsection, we perform POOS forecasting using volatile 1-year data (i.e., data from Nov. 2019 to Oct. 2020). We tactically choose this period data to challenge our new methods for checking whether it can self-adapt to the structural incoherence between pre- and post-pandemic, and we also want to compare our new methods with the existing GE-NoVaS method. For observing affects of pandemic, we can take the price of SP500 index as an example. From \cref{6f1}, it is clearly that the price grew slowly during the normal period form Jan. 2017 to Dec. 2017. However, during the most recent one year, the price fluctuated severely due to the pandemic. 

Similarly, we focus on evaluating the performance of NoVaS-type methods on handling volatile data by doing comparisons with the GARCH-direct method. For executing a comprehensive analysis, we again investigate different methods' performance on stock, index and currency data. 

\subsubsection{Stock data}\label{sssec:stock}
\noindent The POOS forecasting results of volatile 1-year stock datasets are presented in \cref{6t4}. NoVaS-type methods dominate the GARCH-direct method. The performance of the GARCH-direct method is terrible especially for the Bitcoin case. Apart from this overall advantage of NoVaS-type methods, there is no doubt that the GA-NoVaS method manifests greater prediction results than the GE-NoVaS method since it occupies 13 out 27 optimal choices and stands at least 10$\%$ improvement for 5 cases. The parsimonious GA-NoVaS-without-$\beta$ also shows better results than the GE-NoVaS method. This phenomenon lends strong evidence to support our postulation that the GA-NoVaS method is more appropriate to handle volatile data. 

\begin{table}[htbp]
  \caption{Comparison results of using volatile 1-year stock data}
  \label{6t4}
    %\begin{adjustbox}{width=1\textwidth}
    \centering
\scriptsize
\begin{tabular}{lcccccccc}
  \toprule 
 & \thead{\scriptsize GE-NoVaS}  & \thead{\scriptsize GA-NoVaS} & \thead{\scriptsize GA-NoVaS-without-$\beta$}  & \thead{\scriptsize GARCH-direct} \\ 
  \midrule
    NKE-1step & 0.63568  & \textbf{0.63209} & 0.65594 & 1.00000 \\ [1.2pt]
    NKE-5steps & 0.20171  & \textbf{0.19089} & 0.22226  & 1.00000 \\ [1.2pt]
    NKE-30steps & 0.00411  & \textbf{0.00278*} & 0.00340  & 1.00000\\[1.2pt]
    AMZN-1step & 0.97099 & 0.96719 & \textbf{0.90487} &  1.00000 \\[1.2pt]
    AMZN-5steps & 0.88705  & 0.88274 & \textbf{0.72850} &  1.00000\\[1.2pt]
    AMZN-30steps & 0.58124 & 0.62863 & \textbf{0.53310} & 1.00000\\[1.2pt]
    IBM-1step & 0.80222  & 0.79823 & \textbf{0.79509} &  1.00000 \\[1.2pt]
    IBM-5steps & 0.38933 & \textbf{0.37346} & 0.38413 &  1.00000 \\[1.2pt]
    IBM-30steps & 0.01143  & 0.00996\textbf{*} & \textbf{0.00879} &  1.00000 \\[1.2pt]
    MSFT-1step & 0.80133 & \textbf{0.79528} & 0.81582 &  1.00000 \\[1.2pt]
    MSFT-5steps & 0.35567  & \textbf{0.33419} & 0.38022 &  1.00000 \\[1.2pt]
    MSFT-30steps & 0.01342  & 0.01031\textbf{*} & \textbf{0.00784} &  1.00000 \\[1.2pt]
    SBUX-1step & 0.68206  & 0.67067 & \textbf{0.66743} &  1.00000 \\[1.2pt]
    SBUX-5steps & 0.24255  & \textbf{0.23072} & 0.26856 &  1.00000 \\[1.2pt]
    SBUX-30steps & 0.00499  & 0.00337\textbf{*} & \textbf{0.00236} &  1.00000 \\[1.2pt]
    KO-1step & 0.77906  & \textbf{0.75389} & 0.77035 &  1.00000 \\[1.2pt]
    KO-5steps & 0.34941  & \textbf{0.32459} & 0.33405 & 1.00000 \\[1.2pt]
    KO-30steps & 0.01820  & 0.01848 & \textbf{0.01582} &  1.00000 \\[1.2pt]
    MCD-1step & 0.51755  & \textbf{0.51351} & 0.56414 &  1.00000 \\[1.2pt]
    MCD-5steps & 0.10725  & \textbf{0.09714} & 0.17439 &  1.00000 \\[1.2pt]
    MCD-30steps & 3.32E-05  & 2.97E-05\textbf{*} & \textbf{7.62E-06} &  1.00000 \\[1.2pt]
    Tesla-1step & 0.90712  & 0.90250 & \textbf{0.88782} & 1.00000 \\[1.2pt]
    Tesla-5steps & 0.68450  & 0.67935 & \textbf{0.66937} &  1.00000 \\[1.2pt]
    Tesla-30steps & \textbf{0.21643}  & 0.21718 & 0.22395 &  1.00000 \\[1.2pt]
    Bitcoin-1step & 0.36323  & \textbf{0.36260} & 0.36326 & 1.00000 \\[1.2pt]
    Bitcoin-5steps & \textbf{0.01319}  & 0.01321 & 0.01322 &  1.00000 \\[1.2pt]
    Bitcoin-30steps & 7.75E-17 & \textbf{7.65E-17} & 7.75E-17 & 1.00000 \\[1.2pt]
       \bottomrule
    \end{tabular}
    %\end{adjustbox}
            \\
            \raggedright
    \tiny \textit{Note:} The benchmark is the GARCH-direct method, so numerical values in the table corresponding to GARCH-direct method are 1. Other numerical values are relative values compared to the GARCH-direct method. The bold value means that the corresponding method is the optimal choice for this data case. Cell with $*$ means the GA-NoVaS method is at least 10$\%$ better than the GE-NoVaS method or inversely the GE-NoVaS method is at least 10$\%$ better.
\end{table}%

\subsubsection{Currency data}\label{sssec:currency}
\noindent The POOS forecasting results of most recent 1-year currency datasets are presented in \cref{6t5}. One thing should be noticed is that \citet{fryzlewicz2008normalized} implied the ARCH framework seems to be a superior methodology for dealing with the currency exchange data. Therefore, we should not anticipate that GA-NoVaS-type methods can attain much improvement for this data case. However, the GA-NoVaS method still brings off around 26$\%$ and 37$\%$ improvement for 30-steps ahead predictions of CADJPY and CNYJPY, respectively. Besides, the GA-NoVaS-without-$\beta$ method also remains great performance. This surprising result can be seen as an evidence to show GA-NoVaS-type methods are robust to model misspecification.

\begin{table}[htbp]
  \caption{Comparison results of using volatile 1-year currency data}
  \label{6t5}
  \centering
   %\begin{adjustbox}{width=1\textwidth}
\scriptsize
\begin{tabular}{lcccccccc}
  \toprule 
 & \thead{\scriptsize GE-NoVaS}  & \thead{\scriptsize GA-NoVaS} & \thead{\scriptsize GA-NoVaS-without-$\beta$} & \thead{\scriptsize GARCH-direct} \\ 
  \midrule
    CADJPY-1step & 0.46940  & \textbf{0.46382} & 0.48367 &  1.00000 \\[1.2pt]
    CADJPY-5steps & 0.11678  & \textbf{0.11620} & 0.14376 &  1.00000 \\[1.2pt]
    CADJPY-30steps & 0.00584  & \textbf{0.00430*} & 0.00482 &  1.00000 \\[1.2pt]
    EURJPY-1step & 0.95093  & \textbf{0.94682} & 0.95133 &  1.00000 \\[1.2pt]
    EURJPY-5steps & 0.76182  & 0.77091 & \textbf{0.75636} &  1.00000 \\[1.2pt]
    EURJPY-30steps & \textbf{0.16202}  & 0.17956 & 0.18189 & 1.00000 \\[1.2pt]
    USDCNY-1step & 0.98905  & 0.97861 & \textbf{0.95757} &  1.00000 \\[1.2pt]
    USDCNY-5steps & 0.93182  & 0.92614 & \textbf{0.83523} &  1.00000 \\[1.2pt]
    USDCNY-30steps & 0.57171  & \textbf{0.57100} & 0.60131 &  1.00000 \\[1.2pt]
    GBPJPY-1step & 0.86971  & \textbf{0.86474} & 0.87160 &  1.00000 \\[1.2pt]
    GBPJPY-5steps & 0.49749  & 0.49612 & \textbf{0.48842} &  1.00000 \\[1.2pt]
    GBPJPY-30steps & 0.17058  & \textbf{0.16987} & 0.17262 &  1.00000 \\[1.2pt]
    USDINR-1step & 0.97289  & 0.96829 & \textbf{0.93140} & 1.00000 \\[1.2pt]
    USDINR-5steps & 0.80866  & 0.78008 & \textbf{0.75693} &  1.00000 \\[1.2pt]
    USDINR-30steps & \textbf{0.09725}  & 0.09889 & 0.11380 &  1.00000 \\[1.2pt]
    CNYJPY-1step & 0.77812 &  0.77983 & \textbf{0.74586} &  1.00000 \\[1.2pt]
    CNYJPY-5steps & 0.38875  & 0.38407 & \textbf{0.34839} &  1.00000 \\[1.2pt]
    CNYJPY-30steps & 0.08398  & \textbf{0.05240*} & 0.05444 &  1.00000 \\[1.2pt]
       \bottomrule
    \end{tabular}
    %\end{adjustbox}\\
\end{table}%

\subsubsection{Index data}\label{sssec:index}
\noindent The POOS forecasting results of most recent 1-year index datasets are presented in \cref{6t6}. Consistent with conclusions corresponding to previous two classes of data, NoVaS-type methods still have obviously better performance than the GARCH-direct method. Besides this advantage of NoVaS methods, new methods still govern the existing GE-NoVaS method. In addition to these expected results, we find the GE-NoVaS method is even 14$\%$ worse than the GARCH-direct method for 1-step USDX future case. On the other hand, GA-NoVaS-type methods still keep great performance. This phenomenon also appears in \cref{sssec:simuresultsmoeld1-5,sssc:simudifferent,ssc:simusmall,ssec:realdatanormalperiod2years,ssec:realdatanormalperiod1year}. Beyond this, there are 12 cases that the GA-NoVaS method renders more than 10$\%$ improvement compared to the GE-NoVaS method. A most significant case is the 30-steps ahead prediction of Bovespa data where around 60$\%$ improvement is introduced by the GA-NoVaS method compared with GE-NoVaS method. 
\begin{table}[htbp]
  \caption{Comparison results of using volatile 1-year index data}
  \label{6t6}
  \centering
    %\begin{adjustbox}{width=1\textwidth}
\scriptsize
\begin{tabular}{lcccccccc}
  \toprule 
 & \thead{\scriptsize GE-NoVaS}  & \thead{\scriptsize GA-NoVaS} & \thead{\scriptsize GA-NoVaS-without-$\beta$} & \thead{\scriptsize GARCH-direct}  \\ 
  \midrule
    SP500-1step & 0.97294 & 0.95881 & \textbf{0.92854} &  1.00000 \\[1.2pt]
    SP500-5steps & 0.96590  & 0.94457 & \textbf{0.77060} & 1.00000 \\[1.2pt]
    SP500-30steps & 0.34357  & 0.34561 & \textbf{0.30115} &  1.00000 \\[1.2pt]
    Nasdaq-1step & 0.71380  & \textbf{0.70589} & 0.77753 &  1.00000 \\[1.2pt]
    Nasdaq-5steps & 0.29332 & \textbf{0.27007} & 0.36428 &  1.00000 \\[1.2pt]
    Nasdaq-30steps & 0.01223  & \textbf{0.00618*} & 0.00696 &  1.00000 \\[1.2pt]
    NYSE-1step & 0.55741  & 0.55548 & \textbf{0.54598} &  1.00000 \\[1.2pt]
    NYSE-5steps & 0.08994 &  \textbf{0.07666*} & 0.07798 &  1.00000 \\[1.2pt]
    NYSE-30steps & 1.36E-05  & 9.06E-06\textbf{*} & \textbf{6.57E-06} & 1.00000 \\[1.2pt]
    Smallcap-1step & 0.58170  & \textbf{0.57392} & 0.57773 &  1.00000 \\[1.2pt]
    Smallcap-5steps & 0.10270  & 0.10135 & \textbf{0.09628} &  1.00000 \\[1.2pt]
    Smallcap-30steps & 7.00E-05  & 4.33E-05\textbf{*} & \textbf{3.65E-05} &  1.00000 \\[1.2pt]
    BSE-1step & 0.39493  & \textbf{0.37991} & 0.39851 &  1.00000 \\[1.2pt]
    BSE-5steps & 0.03320  & \textbf{0.02829*} & 0.04170 &  1.00000 \\[1.2pt]
    BSE-30steps & 2.45E-05  & 2.19E-05\textbf{*} & \textbf{1.73E-05} &  1.00000 \\[1.2pt]
    DAX-1step & \textbf{0.65372}  & 0.65663 & 0.66097 &  1.00000 \\[1.2pt]
    DAX-5steps & 0.10997  & \textbf{0.10828} & 0.11085 & 1.00000 \\[1.2pt]
    DAX-30steps & 4.97E-05 & \textbf{4.87E-05} & 7.81E-05 & 1.00000\\[1.2pt]
    USDX future-1step & 1.14621  & 1.00926\textbf{*} & 1.03693 &  \textbf{1.00000} \\[1.2pt]
    USDX future-5steps & 0.61075  & 0.53834\textbf{*} & \textbf{0.51997} &  1.00000 \\[1.2pt]
    USDX future-30steps & 0.10723  & \textbf{0.09911} & 0.10063 &  1.00000 \\[1.2pt]
    Bovespa-1step & 0.60031 &  \textbf{0.57316} & 0.60656 &  1.00000 \\[1.2pt]
    Bovespa-5steps & 0.08603 &  \textbf{0.06201*} & 0.09395 & 1.00000 \\[1.2pt]
    Bovespa-30steps & 6.87E-06  & \textbf{2.82E-06*} & 3.19E-06 & 1.00000 \\[1.2pt]
    Djones-1step & 0.56357  & 0.55020 & \textbf{0.54422} &  1.00000 \\[1.2pt]
    Djones-5steps & 0.09810  & \textbf{0.08239*} & 0.08698 &  1.00000 \\[1.2pt]
    Djones-30steps & 4.32E-05  & \textbf{2.22E-05*} & 2.65E-05 & 1.00000 \\[1.2pt]
     BIST-1step & 0.94794  & 0.95313 & \textbf{0.92418} &  1.00000 \\[1.2pt]
     BIST-5steps & \textbf{0.48460}  & 0.49098 & 0.49279 &  1.00000 \\[1.2pt]
     BIST-30steps & \textbf{0.05478} & 0.05980 & 0.05671 & 1.00000 \\[1.2pt]
       \bottomrule
    \end{tabular}
    %\end{adjustbox}\\
\end{table}

\subsection{Summary of real-world data analysis}\label{ssec:summaryofrealdataanalysis}
\noindent After performing extensive real-world data analysis, we can conclude that NoVaS-type methods have generally better performance than the GARCH-direct method. Sometimes, the long-term prediction of GARCH-direct method is impaired due to accumulated errors. Applying NoVaS-type methods can avoid such issue. In addition to this encouraging result, two new NoVaS methods proposed in this article all have greater performance than the existing GE-NoVaS method, especially for analyzing short and volatile data. The satisfactory performance of NoVaS-type methods on predicting Bitcoin data may also open up the application of using NoVaS-type methods to forecast cryptocurrency data.

\section{Comparison of predictive accuracy}\label{sec:comparisonofpredictive}
As illustrated in \cref{sec:intro}, accurate and robust volatility forecasting is an important focus for econometricians. Typically, volatility of returns can be characterized by GARCH-type models. Then, with the Model-free Prediction Principle being proposed, a more accurate NoVaS method was built to predict volatility. This paper further improves the existing NoVaS method by proposing a new transformation structure in \cref{sec:method}. After performing extensive POOS predictions on different classes of data, we find our new methods achieve better prediction performance than traditional GARCH(1,1) model and the existing GE-NoVaS method. The most successful method is the GA-NoVaS-without-$\beta$ method. 

However, one may still think the victory of our new methods is just caused by using specific sample even new methods show lower prediction error (i.e., calculated by \cref{eq:4.1}) for almost all cases. Therefore, we want to learn whether this victory is statistically
significant. We shall notice that \cite{wu2021boosting} applied CW-tests to show removing-$\beta$ idea is appropriate to refine the GE-NoVaS method. Likewise, we are curious about if this refinement is again reasonable for deriving the GA-NoVaS-without-$\beta$ method from the GA-NoVaS method. In this paper, we focus on the CW-test built by \cite{clark2007approximately}\footnote{See \cite{clark2007approximately} for theoretical details of this test, explaining these details is not in the scope of this paper.} which applied an adjusted Mean Squared Prediction Error (MSPE) statistics to test if parsimonious null model and larger model have equal predictive accuracy, see \cite{dangl2012predictive,kong2011predicting,dai2021predicting} for examples of applying this CW-test.

\subsection{CW-test}
Note that the GA-NoVaS-without-$\beta$ method is a parsimonious method compared with the GA-NoVaS method. The reason of removing the $\beta$ term has been illustrated in \cref{ssecmotivation}. Here, we want to deploy the CW-test to make sure the $\beta$-removing idea is not only empirically adoptable but also statistically reasonable. We take several results from \cref{sec:real data} to run CW-tests. However, it is tricky to apply the CW-test on comparing 5-steps and 30-steps aggregated predictions. In other words, the CW-test result for aggregated predictions is ambiguous. It is hard to explain the meaning of a significant small $p$-value. Does this mean a method outperforms the opposite one for all single-step horizons? Or does this mean the method just achieves better performance at some specific future steps? Therefore, we just consider 1-step ahead prediction horizon and CW-test results are tabulated in \cref{7t1}. 

From \cref{7t1}, under a one-sided 5$\%$ significance level, there is only 1 case out of total 28 cases which rejects the null hypothesis. Besides, we should notice that the CW-test still accepts the null hypothesis for 2018-MSFT and volatile period of MCD even the GA-NoVaS method has a better performance value on these cases. Moreover, the GA-NoVaS-without-$\beta$ is more computationally efficient than the GA-NoVaS method. In summary, the reasonability of removing $\beta$ term is shown again by comparing GA-NoVaS and GA-NoVaS-without-$\beta$ methods. 

\begin{table}[H]
\centering
  \caption{CW-tests on 1-step ahead prediction of GA-NoVaS and GA-NoVaS-without-$\beta$ methods}
  \label{7t1}
  \scriptsize 
\begin{tabular}{lccc}
  \toprule 
 & \thead{\scriptsize P-value} & \thead{\scriptsize GA-NoVaS\\ \scriptsize Performance} & \thead{\scriptsize GA-NoVaS-without-$\beta$ \\ \scriptsize Performance}  \\
  \midrule
  2018-AAPL-1step & 0.99 & 0.92 & 0.89 \\ 
  2019-AAPL-1step & 0.08 & 0.81 & 0.82 \\ 
  2018-BAC-1step & 0.63 & 0.94 & 0.93 \\ 
  2019-BAC-1step & 0.49 & 1.05 & 0.99 \\ 
  2018-TSLA-1step & 0.27 & 0.92 & 0.86 \\ 
  2019-TSLA-1step & 0.22 & 1.02 & 0.99 \\
  2018-MCD-1step & 0.57 & 0.98 & 0.94 \\ 
  2019-MCD-1step & 0.19 & 0.96 & 0.95 \\
  2018-MSFT-1step & 0.17 & 0.91 & 0.96 \\
  2019-MSFT-1step & 0.47 & 1.00 & 0.95 \\ 
  2018-Djones-1step & 0.64 & 0.91 & 0.91 \\ 
  2019-Djones-1step & 0.27 & 0.96 & 0.97 \\ 
  2018-Nasdaq-1step & 0.51 & 0.92 & 0.92 \\ 
  2019-Nasdaq-1step & 0.48 & 0.99 & 0.94 \\ 
  2018-NYSE-1step & 0.31 & 0.93 & 0.97 \\ 
  2019-NYSE-1step & 0.11 & 0.91 & 0.92 \\ 
  2018-SP500-1step & 0.42 & 0.91 & 0.88 \\ 
  2019-SP500-1step & 0.32 & 0.97 & 0.93 \\  
  11.2019$\sim$10.2020-IBM-1step & 0.26 & 0.80 & 0.80 \\ 
  11.2019$\sim$10.2020-KO-1step & 0.01 & 0.75 & 0.77 \\ 
  11.2019$\sim$10.2020-MCD-1step & 0.14 & 0.51 & 0.56 \\ 
  11.2019$\sim$10.2020-SBUX-1step & 0.18 & 0.67 & 0.67 \\ 
  11.2019$\sim$10.2020-CADJPY-1step & 0.07 & 0.46 & 0.48 \\ 
  11.2019$\sim$10.2020-CNYJPY-1step & 0.66 & 0.78 & 0.75 \\ 
  11.2019$\sim$10.2020-USDCNY-1step & 0.36 & 0.98 & 0.96 \\ 
  11.2019$\sim$10.2020-EURJP-1step & 0.19 & 0.95 & 0.95 \\ 
  11.2019$\sim$10.2020-Djones-1step & 0.30 & 0.56 & 0.55 \\ 
  11.2019$\sim$10.2020-SP500-1step & 0.25 & 0.59 & 0.58 \\ 

       \bottomrule
    \end{tabular}\\
      \tiny
      \raggedright
     \textit{Note:} The null hypothesis of the CW-test is that parsimonious and larger models have equal MSPE. The alternative is that the larger model has a smaller MSPE. The performance of GA-NoVaS and GA-NoVaS-without-$\beta$ methods are calculated as we did in \cref{sec:real data}, which are  relative values compared with benchmark method (GARCH-direct method).
\end{table}

\section{Conclusion}\label{sec:conclusion}
\noindent In this paper, we show the current state-of-the-art GE-NoVaS and our proposed new methods can avoid error accumulation problem even when long-step ahead predictions are required. These methods outperform GARCH(1,1) model on predicting either simulated data or real-world data under different forecasting horizons. Moreover, the newly proposed GA-NoVaS method is a more stable structure to handle volatile and short data than the GE-NoVaS method. It can also bring significant improvement when the long-term prediction is desired. Additionally, although we reveal that parsimonious variants of GA-NoVaS and GE-NoVaS indeed possess a same structure, the GA-NoVaS-without-$\beta$ method is still more favorable since the corresponding region of model parameter is more complete by design. In summary, the approach to build the NoVaS transformation through the GARCH(1,1) model is sensible and results in superior GA-NoVaS-type methods.

In the future, we plan to explore the NoVaS method in different directions. Our new methods corroborate that and also open up avenues where one can explore other specific transformation structures. In the financial market, the stock data move together. So it would be exciting to see if one can do Model-free predictions in a multiple time series scenario. In some areas, integer-valued time series has important applications. Thus, adjusting such Model-free predictions to deal with count data is also desired. There are also a lot of scopes in proving statistical validity of such predictions. First, we hope a rigorous and systematic way to compare predictive accuracy of NoVaS-type and standard GARCH method can be built. From a statistical inference point of view, one can also construct prediction intervals for these predictions using bootstrap. Such prediction intervals are well sought in the econometrics literature and some results on asymptotic validity of these can be proved. We can also explore dividing the dataset into test and training in some optimal way and see if that can improve performance of these methods. Additionally, since determining the transformation function involves optimization of unknown coefficients, designing a more efficient and precise algorithm may be a further direction to improve NoVaS-type methods. 
\section{Acknowledgement}\label{sec:ackno}
The first author is thankful to Professor Politis for introduction to the topic and useful discussions. The second author's research is partially supported by NSF-DMS 2124222.

\section{Data Availability Statement}\label{sec:dataav}
We have collected all data presented here from \url{www.investing.com} manually. Then, we transform the closing price data to financial log-returns based on \cref{Eq:4.1}.

\bibliographystyle{spbasic}
\bibliography{aggfore}

\begin{thebibliography}{46}
\providecommand{\natexlab}[1]{#1}
\providecommand{\url}[1]{{#1}}
\providecommand{\urlprefix}{URL }
\expandafter\ifx\csname urlstyle\endcsname\relax
  \providecommand{\doi}[1]{DOI~\discretionary{}{}{}#1}\else
  \providecommand{\doi}{DOI~\discretionary{}{}{}\begingroup
  \urlstyle{rm}\Url}\fi
\providecommand{\eprint}[2][]{\url{#2}}

\bibitem[{Abodunrin et~al.(2020)Abodunrin, Oloye, and
  Adesola}]{abodunrin2020coronavirus}
Abodunrin O, Oloye G, Adesola B (2020) Coronavirus pandemic and its implication
  on global economy. International journal of arts, languages and business
  studies 4

\bibitem[{Andersen and Bollerslev(1998)}]{andersen1998answering}
Andersen TG, Bollerslev T (1998) Answering the skeptics: Yes, standard
  volatility models do provide accurate forecasts. International economic
  review pp 885--905

\bibitem[{Ashiya(2003)}]{ashiya2003directional}
Ashiya M (2003) The directional accuracy of 15-months-ahead forecasts made by
  the imf. Applied Economics Letters 10(6):331--333

\bibitem[{Awartani and Corradi(2005)}]{awartani2005predicting}
Awartani BM, Corradi V (2005) Predicting the volatility of the s\&p-500 stock
  index via garch models: the role of asymmetries. International Journal of
  forecasting 21(1):167--183

\bibitem[{Bansal et~al.(2016)Bansal, Kiku, and Yaron}]{bansal2016risks}
Bansal R, Kiku D, Yaron A (2016) Risks for the long run: Estimation with time
  aggregation. Journal of Monetary Economics 82:52--69

\bibitem[{Bellini and Bottolo(2008)}]{bellini2008misspecification}
Bellini F, Bottolo L (2008) Misspecification and domain issues in fitting garch
  (1, 1) models: a monte carlo investigation. Communications in
  Statistics-Simulation and Computation 38(1):31--45

\bibitem[{Ben~Nasr et~al.(2014)Ben~Nasr, Ajmi, and Gupta}]{ben2014modelling}
Ben~Nasr A, Ajmi AN, Gupta R (2014) Modelling the volatility of the dow jones
  islamic market world index using a fractionally integrated time-varying garch
  (fitvgarch) model. Applied Financial Economics 24(14):993--1004

\bibitem[{Bollerslev(1986)}]{bollerslev1986generalized}
Bollerslev T (1986) Generalized autoregressive conditional heteroskedasticity.
  Journal of econometrics 31(3):307--327

\bibitem[{Breitung and Hafner(2016)}]{breitung2016simple}
Breitung J, Hafner CM (2016) A simple model for now-casting volatility series.
  International Journal of Forecasting 32(4):1247--1255

\bibitem[{Chen et~al.(2012)Chen, Yu, and Zivot}]{chen2012predicting}
Chen CH, Yu WC, Zivot E (2012) Predicting stock volatility using after-hours
  information: evidence from the nasdaq actively traded stocks. International
  Journal of Forecasting 28(2):366--383

\bibitem[{Chen(2018)}]{chen2018prediction}
Chen J (2018) Prediction in time series models and model-free inference with a
  specialization in financial return data. PhD thesis, UC San Diego

\bibitem[{Chen and Politis(2019)}]{chen2019optimal}
Chen J, Politis DN (2019) Optimal multi-step-ahead prediction of arch/garch
  models and novas transformation. Econometrics 7(3):34

\bibitem[{Chen and Politis(2020)}]{chen2020time}
Chen J, Politis DN (2020) Time-varying novas versus garch: Point prediction,
  volatility estimation and prediction intervals. Journal of Time Series
  Econometrics 1(ahead-of-print)

\bibitem[{Chortareas et~al.(2011)Chortareas, Jiang, and
  Nankervis}]{chortareas2011forecasting}
Chortareas G, Jiang Y, Nankervis JC (2011) Forecasting exchange rate volatility
  using high-frequency data: Is the euro different? International Journal of
  Forecasting 27(4):1089--1107

\bibitem[{Chud{\`y} et~al.(2020)Chud{\`y}, Karmakar, and Wu}]{chudy2020long}
Chud{\`y} M, Karmakar S, Wu WB (2020) Long-term prediction intervals of
  economic time series. Empirical Economics 58(1):191--222

\bibitem[{Clark and West(2007)}]{clark2007approximately}
Clark TE, West KD (2007) Approximately normal tests for equal predictive
  accuracy in nested models. Journal of econometrics 138(1):291--311

\bibitem[{Dai and Chang(2021)}]{dai2021predicting}
Dai Z, Chang X (2021) Predicting stock return with economic constraint: Can
  interquartile range truncate the outliers? Mathematical Problems in
  Engineering 2021

\bibitem[{Dangl and Halling(2012)}]{dangl2012predictive}
Dangl T, Halling M (2012) Predictive regressions with time-varying
  coefficients. Journal of Financial Economics 106(1):157--181

\bibitem[{Du and Budescu(2007)}]{du2007does}
Du N, Budescu DV (2007) Does past volatility affect investors' price forecasts
  and confidence judgements? International Journal of Forecasting
  23(3):497--511

\bibitem[{Engle and Patton(2001)}]{articleengle2001}
Engle R, Patton A (2001) What good is a volatility model? Quantitative Finance
  1, \doi{10.1088/1469-7688/1/2/305}

\bibitem[{Engle(1982)}]{engle1982autoregressive}
Engle RF (1982) Autoregressive conditional heteroscedasticity with estimates of
  the variance of united kingdom inflation. Econometrica: Journal of the
  Econometric Society pp 987--1007

\bibitem[{Fang et~al.(2018)Fang, Chen, Yu, and Qian}]{fang2018importance}
Fang L, Chen B, Yu H, Qian Y (2018) The importance of global economic policy
  uncertainty in predicting gold futures market volatility: A garch-midas
  approach. Journal of Futures Markets 38(3):413--422

\bibitem[{Fernandes(2020)}]{fernandes2020economic}
Fernandes N (2020) Economic effects of coronavirus outbreak (covid-19) on the
  world economy. Available at SSRN 3557504

\bibitem[{Fiszeder and Perczak(2016)}]{fiszeder2016low}
Fiszeder P, Perczak G (2016) Low and high prices can improve volatility
  forecasts during periods of turmoil. International Journal of Forecasting
  32(2):398--410

\bibitem[{Fryzlewicz et~al.(2008)Fryzlewicz, Sapatinas, Rao
  et~al.}]{fryzlewicz2008normalized}
Fryzlewicz P, Sapatinas T, Rao SS, et~al. (2008) Normalized least-squares
  estimation in time-varying arch models. The Annals of Statistics
  36(2):742--786

\bibitem[{Gonz{\'a}lez-Rivera et~al.(2004)Gonz{\'a}lez-Rivera, Lee, and
  Mishra}]{gonzalez2004forecasting}
Gonz{\'a}lez-Rivera G, Lee TH, Mishra S (2004) Forecasting volatility: A
  reality check based on option pricing, utility function, value-at-risk, and
  predictive likelihood. International Journal of forecasting 20(4):629--645

\bibitem[{Gulay and Emec(2018)}]{gulay2018comparison}
Gulay E, Emec H (2018) Comparison of forecasting performances: Does
  normalization and variance stabilization method beat garch (1, 1)-type
  models? empirical evidence from the stock markets. Journal of Forecasting
  37(2):133--150

\bibitem[{Herrera et~al.(2018)Herrera, Hu, and Pastor}]{herrera2018forecasting}
Herrera AM, Hu L, Pastor D (2018) Forecasting crude oil price volatility.
  International Journal of Forecasting 34(4):622--635

\bibitem[{Karmakar and Roy(2021)}]{karmakar2020bayesian}
Karmakar S, Roy A (2021) Bayesian modelling of time-varying conditional
  heteroscedasticity. Bayesian Analysis 1(1):1--29

\bibitem[{Karmakar et~al.(2020)Karmakar, Chudy, and Wu}]{karmakar2020long}
Karmakar S, Chudy M, Wu WB (2020) Long-term prediction intervals with many
  covariates. To appear at Journal of Time-series Analysis arXiv preprint
  arXiv:201208223

\bibitem[{Kim et~al.(2011)Kim, Rachev, Bianchi, Mitov, and
  Fabozzi}]{kim2011time}
Kim YS, Rachev ST, Bianchi ML, Mitov I, Fabozzi FJ (2011) Time series analysis
  for financial market meltdowns. Journal of Banking \& Finance
  35(8):1879--1891

\bibitem[{Kitsul and Wright(2013)}]{kitsul2013economics}
Kitsul Y, Wright JH (2013) The economics of options-implied inflation
  probability density functions. Journal of Financial Economics 110(3):696--711

\bibitem[{Kong et~al.(2011)Kong, Rapach, Strauss, and
  Zhou}]{kong2011predicting}
Kong A, Rapach DE, Strauss JK, Zhou G (2011) Predicting market components out
  of sample: asset allocation implications. The Journal of Portfolio Management
  37(4):29--41

\bibitem[{Lim and Sek(2013)}]{lim2013comparing}
Lim CM, Sek SK (2013) Comparing the performances of garch-type models in
  capturing the stock market volatility in malaysia. Procedia Economics and
  Finance 5:478--487

\bibitem[{Morikawa(2019)}]{morikawa2019uncertainty}
Morikawa M (2019) Uncertainty in long-term macroeconomic forecasts: Ex post
  evaluation of forecasts by economics researchers. Tech. rep., Research
  Institute of Economy, Trade and Industry (RIETI)

\bibitem[{Olubusoye et~al.(2016)Olubusoye, Yaya, and
  Ojo}]{olubusoye2016misspecification}
Olubusoye OE, Yaya OS, Ojo OO (2016) Misspecification of variants of
  autoregressive garch models and effect on in-sample forecasting. Journal of
  Modern Applied Statistical Methods 15(2):22

\bibitem[{Peters(2001)}]{peters2001estimating}
Peters JP (2001) Estimating and forecasting volatility of stock indices using
  asymmetric garch models and (skewed) student-t densities. Preprint,
  University of Liege, Belgium 3:19--34

\bibitem[{Politis(2003)}]{politis2003normalizing}
Politis DN (2003) A normalizing and variance-stabilizing transformation for
  financial time series. Elsevier Inc.

\bibitem[{Politis(2007)}]{politis2007model}
Politis DN (2007) Model-free versus model-based volatility prediction. Journal
  of Financial Econometrics 5(3):358--359

\bibitem[{Politis(2015)}]{politis2015modelfreepredictionprinciple}
Politis DN (2015) The model-free prediction principle. In: Model-Free
  Prediction and Regression, Springer, pp 13--30

\bibitem[{Politis and McElroy(2019)}]{politis2019time}
Politis DN, McElroy TS (2019) Time series: A first course with bootstrap
  starter. CRC Press

\bibitem[{Taylor(2004)}]{taylor2004volatility}
Taylor JW (2004) Volatility forecasting with smooth transition exponential
  smoothing. International Journal of Forecasting 20(2):273--286

\bibitem[{Wilhelmsson(2006)}]{wilhelmsson2006garch}
Wilhelmsson A (2006) Garch forecasting performance under different distribution
  assumptions. Journal of Forecasting 25(8):561--578

\bibitem[{Wu and Karmakar(2021)}]{wu2021boosting}
Wu K, Karmakar S (2021) Model-free time-aggregated predictions for econometric
  datasets. To appear at Forecasting arXiv preprint arXiv:210102273v3

\bibitem[{Yueh(2020)}]{yueh2020social}
Yueh LY (2020) Social distancing and productivity: how to manage a volatile
  period of growth for the uk economy. British Politics and Policy at LSE

\bibitem[{Zheng(2012)}]{zheng2012empirical}
Zheng X (2012) Empirical analysis of stock return distributions impact upon
  market volatility: Experiences from australia. International Review of
  Business Research Papers 8:156--175

\end{thebibliography}

\newpage
%\appendix
%\section{Comparisons of two parsimonious %variants}\label{appendix:a}
\begin{appendices}
\section{Comparisons of two parsimonious variants}\label{appendix:a}
We asserted that the GA-NoVaS-without-$\beta$ method works better than the GE-NoVaS-without-$\beta$ in \cref{ssec:connection}. Although these two parsimonious variants of GE-NoVaS and GA-NoVaS have a same structure, we showed that regions of their parameters are different. The GA-NoVaS-without-$\beta$ method has a wider parameter space, and this property implies that it is a more complete technique. For substantiating this idea, we compare the forecasting performance of these two parsimonious methods and present results in \cref{tab:ap1}. We can find most of cases are accompanied by very small relative values, which indicates that these two methods stand almost same performance and is in harmony with the fact that they share a same structure. However, we can find there are 21 cases where the GA-NoVaS-without-$\beta$ method works at least 10$\%$ better than the GE-NoVaS-without-$\beta$ method. On the other hand, there is only 8 cases where the GE-NoVaS-without-$\beta$ method shows significantly better results. We shall notice that the GA-NoVaS-without-$\beta$ method is optimized by determining parameters from several grids of values. Therefore, we can imagine the performance of this method will further increase if more refined grids are used. In other words, we can anticipate that the GA-NoVaS-without-$\beta$ method will dominate the GE-NoVaS-without-$\beta$ with subtle grids. 
% Table generated by Excel2LaTeX from sheet 'Sheet3'
\begin{table}
\footnotesize
  \centering
  \caption{Comparisons of parsimonious GE-NoVaS and GA-NoVaS methods}
  \begin{adjustbox}{width=1.5\textwidth}
\begin{tabular}{lccclccclccc}
    \toprule
          & \multicolumn{1}{l}{P-GE} & \multicolumn{1}{l}{P-GA} & \multicolumn{1}{l}{Relative value} &       & \multicolumn{1}{l}{P-GE} & \multicolumn{1}{l}{P-GA} & \multicolumn{1}{l}{Relative value} &       & \multicolumn{1}{l}{P-GE} & \multicolumn{1}{l}{P-GA} & \multicolumn{1}{l}{Relative value} \\
    \midrule
    \multicolumn{4}{c}{500 simulated data} &     AAPL19-5steps  & 0.68191 & 0.68792 & 0.01  & \multicolumn{1}{l}{    AMZN-5steps } & 0.71789 & 0.72850 & 0.01 \\
\cmidrule(lr){1-4}    M1-1step  & 0.85483 & 0.84138 & 0.02  &     AAPL19-30steps  & 0.73823 & 0.72892 & 0.01  & \multicolumn{1}{l}{    AMZN-30steps } & 0.53460 & 0.53310 & 0.00 \\
        M1-5steps  & 0.40510 & 0.40137 & 0.01  &     BAC18-1step  & 0.93031 & 0.92619 & 0.00  & \multicolumn{1}{l}{    IBM-1step } & 0.80744 & 0.79509 & 0.02 \\
        M1-30steps  & 0.03293 & 0.03290 & 0.00  &     BAC18-5steps  & 0.86387 & \textbf{0.72512} & 0.16  & \multicolumn{1}{l}{    IBM-5steps } & 0.40743 & 0.38413 & 0.06 \\
        M2-1step  & 0.98431 & 0.99658 & 0.01  &     BAC18-30steps  & 0.81429 & 0.87630 & 0.07  & \multicolumn{1}{l}{    IBM-30steps } & 0.00918 & 0.00879 & 0.04 \\
        M2-5steps  & 0.90156 & 0.91980 & 0.02  &     BAC19-1step  & 0.97757 & 0.98605 & 0.01  & \multicolumn{1}{l}{    MSFT-1step } & 0.84502 & 0.81582 & 0.03 \\
        M2-30steps  & 0.49761 & 0.48396 & 0.03  &     BAC19-5steps  & 0.89571 & 0.95436 & 0.06  & \multicolumn{1}{l}{    MSFT-5steps } & 0.37528 & 0.38022 & 0.01 \\
        M3-1step  & 0.99360 & 0.98407 & 0.01  &     BAC19-30steps  & 1.01175 & 1.03482 & 0.02  & \multicolumn{1}{l}{    MSFT-30steps } & 0.00732 & 0.00784 & 0.07 \\
        M3-5steps  & 0.94456 & 0.94073 & 0.00  &     MSFT18-1step  & 0.94507 & 0.95921 & 0.01  & \multicolumn{1}{l}{    SBUX-1step } & 0.69943 & 0.66743 & 0.05 \\
        M3-30steps  & 0.92128 & 0.90855 & 0.01  &     MSFT18-5steps  & 0.76646 & 0.74237 & 0.03  & \multicolumn{1}{l}{    SBUX-5steps } & 0.30528 & \textbf{0.26856} & 0.12 \\
        M4-1step  & 0.97484 & 0.99640 & 0.02  &     MSFT18-30steps  & 0.68741 & 0.64717 & 0.06  & \multicolumn{1}{l}{    SBUX-30steps } & 0.00289 & \textbf{0.00236} & 0.18 \\
        M4-5steps  & 0.90862 & 0.95338 & 0.05  &     MSFT19-1step  & 0.98469 & 0.95347 & 0.03  & \multicolumn{1}{l}{    KO-1step } & 0.81309 & 0.77035 & 0.05 \\
        M4-30steps  & 0.68266 & 0.67594 & 0.01  &     MSFT19-5steps  & 1.02387 & 0.95417 & 0.07  & \multicolumn{1}{l}{    KO-5steps } & 0.39679 & \textbf{0.33405} & 0.16 \\
        M5-1step  & 0.96936 & 0.97151 & 0.00  &     MSFT19-30steps  & 0.97585 & 0.98476 & 0.01  & \multicolumn{1}{l}{    KO-30steps } & 0.01963 & \textbf{0.01582} & 0.19 \\
        M5-5steps  & 0.82810 & 0.82747 & 0.00  &     Tesla18-1step  & 0.95885 & \textbf{0.86238} & 0.10  & \multicolumn{1}{l}{    MCD-1step } & 0.58018 & 0.56414 & 0.03 \\
        M5-30steps  & 0.53026 & \textbf{0.47311} & 0.11  &     Tesla18-5steps  & 1.02036 & 1.01560 & 0.00  & \multicolumn{1}{l}{    MCD-5steps } & 0.17887 & 0.17439 & 0.03 \\
        M6-1step  & 0.90196 & 0.93509 & 0.04  &     Tesla18-30steps  & 1.22256 & 1.25005 & 0.02  & \multicolumn{1}{l}{    MCD-30steps } & 0.00001 & 0.00001 & 0.02 \\
        M6-5steps  & 0.75922 & 0.80311 & 0.05  &     Tesla19-1step  & 0.98646 & 0.98955 & 0.00  & \multicolumn{1}{l}{    Tesla-1step } & 0.89253 & 0.88782 & 0.01 \\
        M6-30steps  & 0.40790 & 0.41112 & 0.01  &     Tesla19-5steps  & 0.97523 & 0.96107 & 0.01  & \multicolumn{1}{l}{    Tesla-5steps } & 0.66177 & 0.66937 & 0.01 \\
        M7-1step  & 0.96081 & 0.95932 & 0.00  &     Tesla19-30steps  & 0.87158 & 0.84366 & 0.03  & \multicolumn{1}{l}{    Tesla-30steps } & 0.22460 & 0.22395 & 0.00 \\
        M7-5steps  & 0.84889 & 0.85127 & 0.00  &     Bitcoin18-1step  & 0.99967 & 0.97734 & 0.02  & \multicolumn{1}{l}{    Bitcoin-1step } & 0.36346 & 0.36326 & 0.00 \\
        M7-30steps  & 0.69698 & 0.71391 & 0.02  &     Bitcoin18-5steps  & 1.02008 & 0.93826 & 0.08  & \multicolumn{1}{l}{    Bitcoin-5steps } & 0.01321 & 0.01322 & 0.00 \\
        M8-1step  & 0.92557 & 0.93452 & 0.01  &     Bitcoin18-30steps  & 0.95020 & 0.91364 & 0.04  & \multicolumn{1}{l}{    Bitcoin-30steps } & 0.00000 & 0.00000 & 0.00 \\
\cmidrule(lr){9-12}        M8-5steps  & 0.92549 & 0.93178 & 0.01  &     Bitcoin19-1step  & 0.86795 & 0.87256 & 0.01  & \multicolumn{4}{c}{Volatile 1-year currency data} \\
\cmidrule(lr){9-12}        M8-30steps  & 0.33834 & 0.33853 & 0.00  &     Bitcoin19-5steps  & 0.55620 & 0.56789 & 0.02  & \multicolumn{1}{l}{CADJPY-1step } & 0.48712 & 0.48367 & 0.01 \\
\cmidrule(lr){1-4}    \multicolumn{4}{c}{250 simulated data} &     Bitcoin19-30steps  & 0.00624 & 0.00631 & 0.01  & \multicolumn{1}{l}{    CADJPY-5steps } & 0.13549 & 0.14376 & 0.06 \\
\cmidrule(lr){1-8}    M1-1step  & 0.83168 & 0.83034 & 0.00  & \multicolumn{4}{c}{1-year index data} & \multicolumn{1}{l}{    CADJPY-30steps } & \textbf{0.00394} & 0.00482 & 0.18 \\
\cmidrule(lr){5-8}        M1-5steps  & 0.43772 & 0.43247 & 0.01  & Nasdaq18-1step  & 0.94837 & 0.92421 & 0.03  & \multicolumn{1}{l}{    EURJPY-1step } & 0.94206 & 0.95133 & 0.01 \\
        M1-30steps  & 0.22659 & 0.23035 & 0.02  &     Nasdaq18-5steps  & 0.83585 & 0.78823 & 0.06  & \multicolumn{1}{l}{    EURJPY-5steps } & 0.76727 & 0.75636 & 0.01 \\
        M2-1step  & 0.88781 & 0.87614 & 0.01  &     Nasdaq18-30steps  & 0.74660 & \textbf{0.67389} & 0.10  & \multicolumn{1}{l}{    EURJPY-30steps } & \textbf{0.15350} & 0.18189 & 0.16 \\
        M2-5steps  & 0.52872 & 0.51712 & 0.02  &     Nasdaq19-1step  & 0.93558 & 0.93843 & 0.00  & \multicolumn{1}{l}{    USDCNY-1step } & 0.96766 & 0.95757 & 0.01 \\
        M2-30steps  & 0.73604 & 0.75251 & 0.02  &     Nasdaq19-5steps  & 0.84459 & 0.84051 & 0.00  & \multicolumn{1}{l}{    USDCNY-5steps } & 0.86364 & 0.83523 & 0.03 \\
        M3-1step  & 0.94635 & 0.93693 & 0.01  &     Nasdaq19-30steps  & 0.58924 & 0.59218 & 0.00  & \multicolumn{1}{l}{    USDCNY-30steps } & 0.60121 & 0.60131 & 0.00 \\
        M3-5steps  & 0.96361 & 0.99977 & 0.04  &     NYSE18-1step  & 0.95793 & 0.96619 & 0.01  & \multicolumn{1}{l}{    GBPJPY-1step } & 0.86553 & 0.87160 & 0.01 \\
        M3-30steps  & 0.98872 & 0.99818 & 0.01  &     NYSE18-5steps  & 0.87919 & \textbf{0.75822} & 0.14  & \multicolumn{1}{l}{    GBPJPY-5steps } & 0.49208 & 0.48842 & 0.01 \\
        M4-1step  & 0.92829 & 0.92811 & 0.00  &     NYSE18-30steps  & 0.79466 & \textbf{0.61830} & 0.22  & \multicolumn{1}{l}{    GBPJPY-30steps } & 0.17246 & 0.17262 & 0.00 \\
        M4-5steps  & 0.67482 & 0.67894 & 0.01  &     NYSE19-1step  & 0.90407 & 0.92193 & 0.02  & \multicolumn{1}{l}{    USDINR-1step } & 0.98975 & 0.93140 & 0.06 \\
        M4-30steps  & \textbf{0.71003} & 0.81115 & 0.12  &     NYSE19-5steps  & 0.69822 & 0.71038 & 0.02  & \multicolumn{1}{l}{    USDINR-5steps } & 0.81103 & 0.75693 & 0.07 \\
        M5-1step  & 0.78087 & 0.79075 & 0.01  &     NYSE19-30steps  & 0.18173 & 0.18116 & 0.00  & \multicolumn{1}{l}{    USDINR-30steps } & 0.17966 & \textbf{0.11380} & 0.37 \\
        M5-5steps  & 0.34396 & 0.35155 & 0.02  &     Smallcap18-1step  & 0.91299 & 0.91101 & 0.00  & \multicolumn{1}{l}{    CNYJPY-1step } & 0.79877 & 0.74586 & 0.07 \\
        M5-30steps  & 0.00201 & 0.00194 & 0.03  &     Smallcap18-5steps  & 0.73541 & 0.73223 & 0.00  & \multicolumn{1}{l}{    CNYJPY-5steps } & 0.40569 & \textbf{0.34839} & 0.14 \\
        M6-1step  & 0.94661 & 0.93863 & 0.01  &     Samllcap18-30steps  & 0.48461 & 0.46312 & 0.04  & \multicolumn{1}{l}{    CNYJPY-30steps } & 0.06270 & \textbf{0.05444} & 0.13 \\
\cmidrule(lr){9-12}        M6-5steps  & 0.84719 & 0.85851 & 0.01  &     Smallcap19-1step  & 0.98731 & 0.98482 & 0.00  & \multicolumn{4}{c}{Volatile 1-year index data} \\
\cmidrule(lr){9-12}        M6-30steps  & 0.70301 & 0.70420 & 0.00  &     Smallcap19-5steps  & 0.87700 & 0.85490 & 0.03  & \multicolumn{1}{l}{SP500-1step } & 0.92349 & 0.92854 & 0.01 \\
        M7-1step  & \textbf{0.73553} & 0.83216 & 0.12  &     Samllcap19-30steps  & 0.88825 & 0.90360 & 0.02  & \multicolumn{1}{l}{    SP500-5steps } & 0.75183 & 0.77060 & 0.02 \\
        M7-5steps  & 0.46618 & 0.48020 & 0.03  &     Djones18-1step  & 0.84931 & 0.91192 & 0.07  & \multicolumn{1}{l}{    SP500-30steps } & 0.29793 & 0.30115 & 0.01 \\
        M7-30steps  & 0.06479 & 0.06507 & 0.00  &     Djones18-5steps  & 0.86017 & \textbf{0.76226} & 0.11  & \multicolumn{1}{l}{    Nasdaq-1step } & 0.75350 & 0.77753 & 0.03 \\
        M8-1step  & 0.76586 & 0.83340 & 0.08  &     Djones18-30steps  & 0.66418 & \textbf{0.56854} & 0.14  & \multicolumn{1}{l}{    Nasdaq-5steps } & 0.33519 & 0.36428 & 0.08 \\
        M8-5steps  & \textbf{0.38107} & 0.45665 & 0.17  &     Djones19-1step  & 0.96365 & 0.96977 & 0.01  & \multicolumn{1}{l}{    Nasdaq-30steps } & \textbf{0.00599} & 0.00696 & 0.14 \\
        M8-30steps  & 0.00918 & 0.00983 & 0.07  &     Djones19-5steps  & 0.89542 & 0.91238 & 0.02  & \multicolumn{1}{l}{    NYSE-1step } & 0.57174 & 0.54598 & 0.05 \\
\cmidrule(lr){1-4}    \multicolumn{4}{c}{2-years real-world data} &     Djones19-30steps  & 0.80304 & 0.81803 & 0.02  & \multicolumn{1}{l}{    NYSE-5steps } & 0.10182 & \textbf{0.07798} & 0.23 \\
\cmidrule(lr){1-4}     AAPL-1step  & 0.98261 & 0.97836 & 0.00  &     SP50018-1step  & 0.90469 & 0.88405 & 0.02  & \multicolumn{1}{l}{    NYSE-30steps } & 0.00001 & 0.00001 & 0.01 \\
        AAPL-5steps  & 1.00090 & 0.96999 & 0.03  &     SP50018-5steps  & 0.90544 & \textbf{0.75646} & 0.16  & \multicolumn{1}{l}{    Smallcap-1step } & 0.60931 & 0.57773 & 0.05 \\
        AAPL-30steps  & 0.97326 & 0.96174 & 0.01  &     SP50018-30steps  & 0.83210 & \textbf{0.72104} & 0.13  & \multicolumn{1}{l}{    Smallcap-5steps } & 0.10337 & 0.09628 & 0.07 \\
        BAC-1step  & 1.00416 & 1.02780 & 0.02  &     SP50019-1step  & 0.92183 & 0.93162 & 0.01  & \multicolumn{1}{l}{    Smallcap-30steps } & 0.00006 & \textbf{0.00004} & 0.39 \\
        BAC-5steps  & 0.97542 & 0.99399 & 0.02  &     SP50019-5steps  & 0.75579 & 0.77434 & 0.02  & \multicolumn{1}{l}{    BSE-1step } & 0.39745 & 0.39851 & 0.00 \\
        BAC-30steps  & 0.99533 & 1.00363 & 0.01  &     SP50019-30steps  & 0.29796 & 0.30127 & 0.01  & \multicolumn{1}{l}{    BSE-5steps } & 0.04109 & 0.04170 & 0.01 \\
        MSFT-1step  & 0.97376 & 0.96185 & 0.01  &     BSE18-1step  & 0.94676 & 0.92568 & 0.02  & \multicolumn{1}{l}{    BSE-30steps } & 0.00002 & 0.00002 & 0.05 \\
        MSFT-5steps  & 0.96601 & 0.95271 & 0.01  &     BSE18-5steps  & 0.82886 & 0.84408 & 0.02  & \multicolumn{1}{l}{    DAX-1step } & 0.64727 & 0.66097 & 0.02 \\
        MSFT-30steps  & 0.91046 & 0.88515 & 0.03  &     BSE18-30steps  & 0.44818 & 0.44092 & 0.02  & \multicolumn{1}{l}{    DAX-5steps } & 0.10356 & 0.11085 & 0.07 \\
        MCD-1step  & 0.98873 & 0.99005 & 0.00  &     BSE19-1step  & 0.67694 & 0.66667 & 0.02  & \multicolumn{1}{l}{    DAX-30steps } & 0.00007 & 0.00008 & 0.06 \\
        MCD-5steps  & 0.98059 & 0.97777 & 0.00  &     BSE19-5steps  & 0.23665 & 0.23603 & 0.00  & \multicolumn{1}{l}{    USDX future-1step } & 0.99640 & 1.03693 & 0.04 \\
        MCD-30steps  & 0.99512 & 0.99418 & 0.00  &     BSE19-30steps  & 0.02890 & 0.02888 & 0.00  & \multicolumn{1}{l}{    USDX future-5steps } & 0.54834 & 0.51997 & 0.05 \\
\cmidrule(lr){1-4}    \multicolumn{4}{c}{1-year stock data} &     BIST18-1step  & 0.92271 & 0.94138 & 0.02  & \multicolumn{1}{l}{    USDX future-30steps } & 0.10278 & 0.10063 & 0.02 \\
\cmidrule(lr){1-4}    MCD18-1step & 0.95332 & 0.94412 & 0.01  &     BIST18-5steps  & 0.80837 & 0.81417 & 0.01  & \multicolumn{1}{l}{    Bovespa-1step } & 0.57558 & 0.60656 & 0.05 \\
    MCD18-5steps  & 0.88773 & 0.88151 & 0.01  &     BIST18-30steps  & 1.09665 & 1.09900 & 0.00  & \multicolumn{1}{l}{    Bovespa-5steps } & \textbf{0.07447} & 0.09395 & 0.21 \\
        MCD18-30steps  & 0.61097 & 0.61153 & 0.00  &      BIST19-1step  & 0.95467 & 0.98234 & 0.03  & \multicolumn{1}{l}{    Bovespa-30steps } & \multicolumn{1}{l}{\textbf{2.04E-6}} & \multicolumn{1}{l}{3.19E-6} & 0.36 \\
        MCD19-1step  & 0.93141 & 0.94559 & 0.01  &      BIST19-5steps  & 0.82898 & 0.85370 & 0.03  & \multicolumn{1}{l}{    Djones-1step } & 0.57550 & 0.54422 & 0.05 \\
        MCD19-5steps  & 0.90061 & 0.90602 & 0.01  &      BIST19-30steps  & 0.88511 & 0.85743 & 0.03  & \multicolumn{1}{l}{    Djones-5steps } & 0.11554 & \textbf{0.08698} & 0.25 \\
\cmidrule(lr){5-8}        MCD19-30steps  & 0.80805 & 0.77976 & 0.04  & \multicolumn{4}{c}{Volatile 1-year stock data} & \multicolumn{1}{l}{    Djones-30steps } & \textbf{0.00001} & 0.00003 & 0.53 \\
\cmidrule(lr){5-8}        AAPL18-1step  & 0.96311 & 0.89283 & 0.07  & NKE-1step  & 0.65295 & 0.65594 & 0.00  & \multicolumn{1}{l}{     BIST-1step } & 0.90832 & 0.92418 & 0.02 \\
        AAPL18-5steps  & 0.85114 & \textbf{0.71233} & 0.16  &     NKE-5steps  & 0.22550 & 0.22226 & 0.01  & \multicolumn{1}{l}{     BIST-5steps } & 0.47480 & 0.49279 & 0.04 \\
        AAPL18-30steps  & 0.35731 & 0.37081 & 0.04  &     NKE-30steps  & 0.00337 & 0.00340 & 0.01  & \multicolumn{1}{l}{     BIST-30steps } & 0.05550 & 0.05671 & 0.02 \\
        AAPL19-1step  & 0.80948 & 0.81872 & 0.01  &     AMZN-1step  & 0.90200 & 0.90487 & 0.00  &       &       &       &  \\
        \bottomrule
    \end{tabular}%
      \end{adjustbox}
%\begin{adjustbox}{width=1.5\textwidth}
\begin{tabular}{p{1.5\textwidth}}
\scriptsize
\textit{Note:} \textit{P-GE} and \textit{P-GA} columns represent GE-NoVaS-without-$\beta$ and GA-NoVaS-without-$\beta$ methods' relative forecasting performance compared \newline to the GARCH-direct method, respectively. The \textit{Relative value} column measures how much the GA-NoVaS-without-$\beta$ method is better than \newline  the GE-NoVaS-without-$\beta$ method or how much the GE-NoVaS-without-$\beta$ method is better than the GA-NoVaS-without-$\beta$ method, i.e., it is \newline calculated by $(\text{max(P-GE,P-GA)}-\text{min(P-GE,P-GA)})/\text{max(P-GE,P-GA)}$. The bold values mark cases where one of these two methods is at least \newline 10$\%$ better than the another one based on the relative prediction performance.\\
\end{tabular}
%\end{adjustbox}
  \label{tab:ap1}%
  \end{table}

\end{appendices}

\end{document}